\documentclass[12pt,draftcls,onecolumn]{IEEEtran}

\usepackage{nopageno}
\usepackage{psfrag}
\usepackage[final,dvips]{graphicx}
\usepackage{amsmath}
\usepackage{amssymb}
\usepackage{epsfig}
\usepackage{amsthm}
\usepackage[square,numbers,sort&compress]{natbib}
\usepackage[table,xcdraw]{xcolor}
\usepackage{array}
\usepackage{algorithm}
\usepackage{algpseudocode} 
\usepackage{mathtools}
\usepackage[labelformat=simple]{subcaption}
\usepackage{booktabs}

\usepackage{url}

\usepackage[hidelinks]{hyperref}
\usepackage{breakurl}

\DeclareCaptionLabelSeparator{periodspace}{.\quad}
\DeclareGraphicsExtensions{pst,eps}
\usepackage{scalerel}
\usepackage{tikz}
\usetikzlibrary{svg.path}

\definecolor{orcidlogocol}{HTML}{A6CE39}
\tikzset{
  orcidlogo/.pic={
    \fill[orcidlogocol] svg{M256,128c0,70.7-57.3,128-128,128C57.3,256,0,198.7,0,128C0,57.3,57.3,0,128,0C198.7,0,256,57.3,256,128z};
    \fill[white] svg{M86.3,186.2H70.9V79.1h15.4v48.4V186.2z}
                 svg{M108.9,79.1h41.6c39.6,0,57,28.3,57,53.6c0,27.5-21.5,53.6-56.8,53.6h-41.8V79.1z M124.3,172.4h24.5c34.9,0,42.9-26.5,42.9-39.7c0-21.5-13.7-39.7-43.7-39.7h-23.7V172.4z}
                 svg{M88.7,56.8c0,5.5-4.5,10.1-10.1,10.1c-5.6,0-10.1-4.6-10.1-10.1c0-5.6,4.5-10.1,10.1-10.1C84.2,46.7,88.7,51.3,88.7,56.8z};
  }
}

\newcommand\orcidicon[1]{$^{\href{https://orcid.org/#1}{\mbox{\scalerel*{
\begin{tikzpicture}[yscale=-3,xscale=-3,transform shape]
\pic{orcidlogo};
\end{tikzpicture}
}{b}}}}$}

\newcommand{\A}{\boldsymbol{\Lambda}}
\newcommand{\U}{\mathbf{U}}

\newcommand{\Q}{\mathbf{Q}}
\newcommand{\D}{\mathbf{D}}
\newcommand{\Cov}{\boldsymbol{\Sigma}} 

\newcommand{\z}{\mathbf{z}}
\newcommand{\x}{\mathbf{x}}

\newcommand{\h}{\mathbf{h}}
\newcommand{\e}{\mathbf{e}}

\renewcommand{\b}{\mathbf{b}}

\newcommand{\mean}{\boldsymbol{\mu}} 

\newcommand{\Hyp}{\mathcal{H}}
\newcommand{\CN}{\mathcal{C}\mathcal{N}}
\newcommand{\prob}{\mathbb{P}}

\DeclareMathOperator{\argmax}{arg\,max}

\newcommand{\Int}{\int\limits}
\newcommand{\QF}[2][]{\|#2\|_{#1}^2}
\newcommand{\qf}[2][]{#2^\dag #1 #2}
\usepackage{ifthen}

\newcommand{\NRx}{N_{\text{Rx}}}
\newcommand{\pFA}{p_{\text{FA}}}
\newcommand{\pMD}{p_{\text{MD}}}

\newcommand{\KRice}{K_{\text{Rice}}}

\newcommand{\Ka}{N_{\text{RRH}}}
\newcommand{\Na}{N_{\text{a}}}

\newcommand{\DOk}{\Delta\Omega_j}
\newcommand{\SEEk}{S_{\text{EE}}^{(j)}}

\newcommand{\SEAk}{S_{\text{EA}}^{(j)}}

\newcommand{\meanA}{\mean_{\text{A}}}
\newcommand{\meanE}{\mean_{\text{E}}}
\newcommand{\CovE}{\Cov_{\text{E}}}
\newcommand{\CovA}{\Cov_{\text{A}}}

\newcommand{\hE}{\h_{\text{E}}}
\newcommand{\xiE}{\xi_{\text{E}}}
\newcommand{\xiA}{\xi_{\text{A}}}

\newcommand{\QE}{\Q_{\text{E}}}
\newcommand{\QA}{\Q_{\text{A}}}

\newtheoremstyle{mystyle}
  {3pt}
  {3pt}
  {\itshape}
  {10pt}
  {\itshape}
  {:}
  { }
  {}
\theoremstyle{mystyle}
\newtheorem{thm}{Theorem}
\newtheorem{lemma}{Lemma}
\newtheorem{remark}{Remark}
\newtheorem{prop}{Proposition}
\newtheorem{cor}{Corollary}
\newtheorem{definition}{Definition}

\usepackage{todonotes}
\newcounter{todocounter}

\newcounter{notecounter}

\begin{document}
	
\title{\textcolor{black}{Worst-Case Detection Performance for Distributed SIMO Physical Layer Authentication}}
\author{\IEEEauthorblockN{	Henrik Forssell\orcidicon{0000-0003-4961-5973}, \IEEEmembership{Student Member, IEEE},
							Ragnar Thobaben\orcidicon{0000-0001-9307-484X}, \IEEEmembership{Member, IEEE},
							}

\thanks{
This work was supported in part by the Swedish Civil Contingencies Agency, MSB, through the \href{https://www.kth.se/dcs/research/secure-control-systems/cerces/cerces-center-for-resilient-critical-infrastructures-1.609722}{CERCES} project.
The authors want to thank professor James Gross for discussions and insights that contributed to the completion of this work as well as for proof-reading the final manuscript.
H. Forssell and R. Thobaben are with the School of Electrical Engineering and Computer Science, KTH Royal Institute of Technology, Stockholm, Sweden (e-mail: hefo@kth.se; ragnart@kth.se).
	}
}
\maketitle

\begin{abstract}
\textcolor{black}{
Feature-based physical layer authentication (PLA) schemes, using position-specific channel characteristics as identifying features, can provide lightweight protection against impersonation attacks in overhead-limited applications like e.g., mission-critical and low-latency scenarios. However, with PLA-aware attack strategies, an attacker can maximize the probability of successfully impersonating the legitimate devices. In this paper, we provide worst-case detection performance bounds under such strategies for a distributed PLA scheme that is based on the channel-state information (CSI) observed at multiple distributed remote radio-heads. This distributed setup exploits the multiple-channel diversity for enhanced detection performance and mimics distributed antenna architectures considered for 4G and 5G radio access networks. We consider (i) a power manipulation attack, in which a single-antenna attacker adopts optimal transmit power and phase; and (ii) an optimal spatial position attack. Interestingly, our results show that the attacker can achieve close-to-optimal success probability with only statistical CSI, which significantly strengthens the relevance of our results for practical scenarios. Furthermore, our results show that, by distributing antennas to multiple radio-heads, the worst-case missed detection probability can be reduced by 4 orders of magnitude without increasing the total number of antennas, illustrating the superiority of distributed PLA over a co-located antenna setup.
}

\end{abstract}


\vspace{-1ex}
 \section{Introduction}		\label{sec:introduction}

Feature-based physical layer authentication (PLA) of wireless communications is currently researched as a means of providing enhanced security in applications where quick authentication with low complexity and security overhead is desirable.
The basic idea of such schemes is to verify the legitimacy of a message by exploiting characteristic features of the user locations or hardware chipsets that can be inferred from the received PHY-layer signals.
Several different PHY-layer features can be used for PLA, ranging from hardware-specific features such as carrier frequency offsets (CFO)~\cite{Hou2014}, offsets in clock frequencies~\cite{Jana2010}, and switching transients~\cite{Danev2009}, to location-specific features such as the received signal strength indicator (RSSI)~\cite{Hussain2009}, the wide band multi-path channel~\cite{Xiao2007}, and multiple-antenna channels~\cite{Abdelaziz2019,Baracca2012}.
The major advantage of these schemes is that they require no additional security overhead, as opposed to cryptographic authentication and tag-based PLA that rely on embedding a pre-agreed secret key~\cite{Yu2008}.

\textcolor{black}{A general issue with feature-based PLA schemes is the possibility for an attacker to use various smart strategies to mimic the legitimate user features.
For instance, it is well known that CFOs can be impersonated by adapting the transmit frequency to match the legitimate transmitter's, and RSSIs can be altered by manipulating the transmit power.
With PLA based on more diverse channel features, an attacker can exploit ray-tracing and statistical knowledge of the legitimate channel to optimally mimic the legitimate feature~\cite{Baracca2012}, and with location-specific features, an attacker can choose an optimal position for mimicking the legitimate channel.
Such vulnerabilities clearly undermine the trustworthiness of PLA, and the question arises which level of security can be guaranteed under any kind of attack.
Considering that PLA is envisioned to provide security in overhead-limited communications, including mission-critical and ultra-reliable low-latency communications (URLLC)~\cite{Chen2019,Weinand2019} where security breaches can have catastrophic effects, provable security guarantees for these schemes are needed.}

\subsection{Contributions of this Paper}

\textcolor{black}{
In this paper, we provide methods for deriving worst-case detection performance guarantees for feature-based PLA schemes subject to optimally designed single-antenna attacks.
We consider two attack strategies: (i) a power manipulation attack, where the attacker adapts power and phase at a single-antenna transmitter in order to shape its channel response by scaling and phase rotation, and (ii) an optimal position attack, where the attacker chooses the spatial position so as to optimize her success probability.
We derive the worst-case bounds for a distributed PLA scheme which is based on the channel-state information (CSI) vectors observed at multiple distributed reception points.
The CSI statistics are functions of the received power and angle-of-arrival (AoA) and the PLA scheme can therefore be viewed as a form of multiple-array line-of-sight (LOS) receive beam-forming.
In the distributed setup, conceptually, the impersonation task of the attacker becomes increasingly difficult due to the diverse observations of features at the multiple receivers, and our worst-case analysis allow us to quantify the performance gains obtained from the distributed approach..
Moreover, this setup is well motivated by the recent trends in 5G towards exploiting distributed antenna architectures (e.g., coordinated multi-point reception and Cloud RAN) to realize the strict reliability requirements of mission-critical communications~\cite{Panigrahi2017,Alonzo2020}.
Therefore, distributed PLA, paired with these types of worst-case performance guarantees, is a promising solution towards secure mission-critical communications.
}

\textcolor{black}{
In summary, the contributions of this paper are:
}

\begin{itemize}
\item \textcolor{black}{We derive the optimal transmit-power manipulation strategy under perfect channel-state information (CSI) knowledge at the attacker (which corresponds to a worst-case attacker) and the corresponding missed detection probability.
This result, which serves as a worst-case bound for a given attacker location, is derived in closed-form for a single receive radio-head and as a saddle-point approximation for the multiple radio-head case.}
\item \textcolor{black}{We show that the saddle-point approximation can be used to obtain the missed detection probability under any power manipulation strategy, and in particular, for a strategy that only requires statistical CSI knowledge. This greatly extends the practical relevance of our contribution.}
\item \textcolor{black}{We characterize the optimal attacker position with respect to a given network deployment under strong LOS assumptions and provide a heuristic truncated search algorithm that significantly reduces the search-space to a set of locally optimal attack positions.
We show that the truncated search algorithm efficiently finds the optimal attack position, and hence, constitutes a powerful tool for planning, analyzing, and optimizing deployments from a security perspective.}
\end{itemize}







\subsection{\textcolor{black}{Related Work}}

\textcolor{black}{
There are only a few previous works that have considered optimized attack strategies against PLA.
In~\cite{Baracca2012}, the authors consider an attacker that uses a forged channel to optimally attack a MIMO PLA scheme.
In their work, the attacker is assumed to be able to produce any forged channel with respect to the receiver but the practical strategy for achieving this is not discussed.
In a closely related setup, the work in \cite{Ferrante2015} derives the outer region of the achievable detection performance for a MIMO/OFDM-based PLA scheme.
They utilize an information theoretic bound based on the Kullback-Leibler divergence that is optimized over the space of attack distributions.
The performance evaluation in~\cite{Senigagliesi2020} also considers a forged channel, however, as opposed to our work, the PLA scheme in their work is based on a machine-learning approach and they do not provide any closed-form solutions to the missed detection probability under the defined attack strategy.
In comparison to these previous works, the performance bounds we provide in this paper are instead based on closed form solutions for the optimal transmit power and phase at a single-antenna attacker.
Moreover, none of these previous works consider attack strategies against distributed PLA and, thus, our work is the first to quantify the benefits of this approach in terms of worst-case detection performance.}

\textcolor{black}{
The PLA scheme we analyze is based on the generalized likelihood-ratio test (GLRT) and, thus, shares a similar mathematical formulation with several previously proposed PLA schemes based on multi-dimensional complex Gaussian features~\cite{Xiao2007,Xiao2008,Baracca2012}.
Therefore, parts of our results are useful for deriving the worst-case bounds for these schemes as well.
Schemes based on AoAs have previously been proposed for vehicular communications in~\cite{Abdelaziz2016,Abdelaziz2019} and these schemes are based on a similar LOS phased-array model that we employ in this paper.
The work in~\cite{Mahmood2017} is to our knowledge the only previous work that considers PLA in a distributed setting; however, their work focuses on decision fusion based on compressed sensing and does not consider attack strategies.
Similarly, our previous work~\cite{Forssell2019b} was an initial study towards PLA in the distributed setup without considering any attack strategies.
}

\subsection{Paper Outline}

The rest of this paper is organized as follows:
Section~\ref{sec:system_model} introduces the considered system model, authentication scheme, and the problem formulation.
In Section~\ref{sec:opt_power_attack}, we analyze the power manipulation attack and provide the corresponding missed detection probabilities.
In Section~\ref{sec:opt_attack_pos}, we study the optimal attacker positions and define the heuristic optimization approach.
Section~\ref{sec:num_res} provides the numerical evaluation of the derived performance bounds and compares different deployment strategies.
Finally, the paper is concluded in Section~\ref{sec:conclusion}. 

\textit{Notation:} 
Matrices are represented by bold capital symbols $\mathbf{X}$, and $\mathbf{X}^T$ and $\mathbf{X}^\dag$ denote the matrix transpose and conjugate transpose, respectively. 
We let $\mathbf{I}$ denote the identity matrix.
Vectors with entries $x_i$ are represented by bold lower-case symbols $\mathbf{x}$ for which we let $\|\mathbf{x}\|=\sqrt{|x_1|^2+...+|x_n|^2}$ denote the Euclidian norm and $\QF[\mathbf{A}]{\mathbf{x}}=\qf[\mathbf{A}]{\mathbf{x}}$ denote the complex quadratic form.
We use $F_X(x)$ to represent the cumulative distribution function of a random variable $X$. 
We let $\CN(\mean,\Cov)$ represent the multivariate proper complex Gaussian distribution with mean $\mean$ and covariance matrix $\Cov$, $\mathcal{N}(\mean,\Cov)$ the corresponding real-valued Gaussian distribution, $\chi^2_k$ a central $\chi^2$ distribution with $k$ degrees of freedom, and $\chi_k^2(\lambda)$ a non-central $\chi^2$ distribution with $k$ degrees of freedom and non-centrality parameter $\lambda$.

 \vspace{-2ex}
 \section{System Model and Preliminaries}		\label{sec:system_model}
\definecolor{dgreen}{rgb}{0.0, 0.5, 0.0}
\begin{figure}[h]
\centering
\psfrag{A}[l][]{\scriptsize Remote radio-head}
\psfrag{B}[l][]{\scriptsize Feature values/soft PLA decisions}
\psfrag{C}[l][]{\scriptsize Physical exclusion region}
\psfrag{D}[l][]{\scriptsize Worst-case single-antenna attacker}
\psfrag{E}[l][]{\scriptsize Optimal power/phase}
\psfrag{F}[l][]{\scriptsize Optimal position}
\psfrag{G}[l][]{\scriptsize Centralized baseband processor}
\psfrag{b}[][]{\small \textcolor{blue}{Bob}}
\psfrag{a}[][]{\small \textcolor{dgreen}{Alice}}
\psfrag{e}[][]{\small \textcolor{red}{Eve}}
\psfrag{m}[][]{\scriptsize $\h_{\text{A}}^{(1)}$}
\psfrag{n}[][]{\scriptsize $\h_{\text{A}}^{(2)}$}
\psfrag{o}[][]{\scriptsize $\h_{\text{A}}^{(3)}$}
\psfrag{p}[][]{\tiny AoA}
\includegraphics[width=1\textwidth]{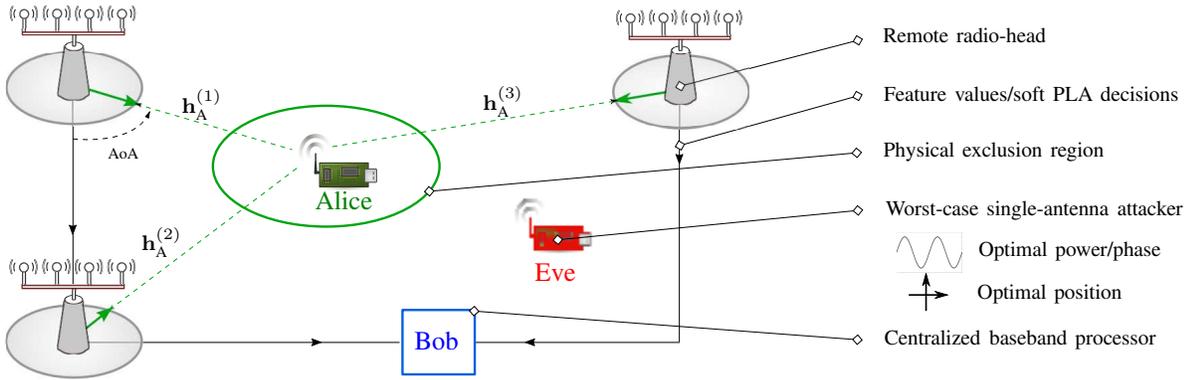}
\vspace{-2ex}
\caption{System deployment consisting of wireless sensors communicating in uplink to multiple-antenna remote radio-heads (RRHs), a centralized baseband processor (Bob), and a worst-case single-antenna adversary (Eve).}
\vspace{-3ex}
\label{fig:system_model}
\end{figure}

In this paper, we analyze authentication of uplink transmissions in a wireless system running a mission-critical application (e.g., sensors sending data to an industrial automation process). 
As depicted in Fig.~\ref{fig:system_model}, we consider a distributed system architecture consisting of $\Ka$ remote radio-heads (RRHs), each equipped with $\NRx$ receive antennas, connected to a centralized baseband processor we refer to as Bob.
There is a legitimate single-antenna transmitter referred to as Alice and a rouge transmitter, referred to as Eve, who is attempting to compromise the system by impersonating Alice.
The PLA scheme considered in this paper, which is formally introduced in Section~\ref{sub:physical_layer_authentication}, is designed to protect the system against Eve's impersonation attempts by comparing the CSI of each transmission against a pre-stored feature bank.
\textcolor{black}{Since the PLA scheme is based on the spatial characteristics of the CSI, Eve is assumed to be positioned outside a physical exclusion region around Alice.}

\vspace{-3ex}
\subsection{\textcolor{black}{Channel Model}} 

The PHY-layer channels from the devices to the RRHs are modeled as narrowband single-input multiple-output (SIMO) channels centered at frequency $f_c$ and subject to Rice fading.
We let $\h_{i}^{(j)}$ denote the $(\NRx\times 1)$ complex channel vector from device $i=\{A,E\}$\footnote{Note that the model presented in this section extends to multiple devices $i=1,\cdots,N_{\text{devices}}$; however, in this paper we only consider the single legitimate device Alice.} to array $j$ and model them as circular-symmetric complex Gaussian (CSCG) vectors $\h_{i}^{(j)}\sim \CN(\mean_i^{(j)},\Cov_i^{(j)})$ (i.e., narrowband SIMO Rice fading), where we let $\mean_i^{(j)}$ and $\Cov_i^{(j)}$ denote the corresponding mean vector and covariance matrix, respectively.
We assume that $\mathbb{E}[\|\h_{i,j}^{(j)}\|^2] = P_i^{(j)}\NRx$ where $P_i^{(j)}$ represents the average power received per antenna and that the covariance matrices take the form $\Cov_i^{(j)} = \frac{P_i^{(j)}}{K_{\text{Rice}}+1}\A$, where $\A$ is a fixed correlation matrix identical for each RRH.
We use a parametric model for the received power per antenna $P_i^{(j)} = \left(\frac{\lambda_c}{4\pi d_i^{(j)}}\right)^\beta P_{\text{Tx},i}$, with $\lambda_c = c/f_c$ being the wavelength, $d_i^{(j)}$ the distance, $P_{\text{Tx},i}$ the transmit power, and $\beta$ a path-loss exponent (i.e., $\beta=2$ represents free-space path loss).
Note also that we deliberately let both transmit power and path-loss be parts of the CSI vectors $\h_{i}^{(j)}$ through $P_i^{(j)}$ since the receiver in practice is unable to differentiate these from each other based on the received signals.

We adopt a phased-array model of the expected value of the channels which then becomes a location-specific statistic of the channel distributions.
We denote by $\Phi_i^{(j)}$ the spatial angle-of-arrival from transmitter $i$ w.r.t. array $j$ and let

\begin{equation}
\mean_i^{(j)} = \sqrt{\frac{P_i^{(j)}  \KRice}{\KRice+1}} \times e^{-\frac{j2\pi d_i^{(j)}}{\lambda_c}}\e(\Omega_i^{(j)}),
\label{eq:phased_array}
\end{equation}
where the array-response vector $\e(\Omega) =\begin{bmatrix} 1, e^{-j2\pi\Delta_r\Omega}, \cdots, e^{-j2\pi\Delta_r(\NRx-1)\Omega} \end{bmatrix}^T$ is modeling the phase differences between antenna elements in terms of the angular sine $\Omega = \sin(\Phi)$ and normalized antenna separation $\Delta_r$.


%

\textcolor{black}{
Next, we define the PLA scheme employed by the centralized receiver Bob.
}

\vspace{-3ex}
\subsection{Physical Layer Authentication Scheme} 
\label{sub:physical_layer_authentication}

In this paper, we consider the authentication problem where Bob receives a message $m$ with uncertainty as to whether it originated from Alice or Eve.
We assume that the message is intercepted by every RRH and we denote by $\tilde{\h}_{m}^{(j)}$ the observed channel-state vector at receive array $j$.
We denote by $\Hyp_0$ the hypothesis that the message is from Alice, i.e., that $\tilde{\h}_{m}^{(j)} =\h_{A}^{(j)}$, and by $\Hyp_1$ the hypothesis that it is from Eve, i.e., that $\tilde{\h}_{m}^{(j)} =\h_{E}^{(j)}$.
In general, $\tilde{\h}_{m}^{(j)}$ would be a channel-state estimate with limited precision; however, to simplify the analysis in the following we assume perfect CSI\footnote{
\textcolor{black}{
Note that parts of the results in this paper can be generalized to imperfect CSI by extending $\Cov_i^{(j)}=\frac{P_i^{(j)}}{K_{\text{Rice}}+1}\A + \sigma_n^2\mathbf{I}$, where $\sigma_n^2$ is the estimation noise variance. However, such analysis is left out due to space limitations.}
}
at the RRHs.



Bob's objective is to centrally decide whether to accept the message or not based on information received from the RRHs.
For that purpose, we construct the $(\NRx\Ka \times 1)$ CSI vector $\tilde{\h}_{m} = [[\tilde{\h}_{m}^{(1)}]^T\cdots [\tilde{\h}_{m}^{(\Ka)}]^T]^T$.
Given that the message is authentic (i.e., $\Hyp_0$ is true), we have $\tilde{\h}_{m}\sim \CN(\meanA,\CovA)$, with

 \vspace{-2ex}\begin{equation}
\meanA = \begin{bmatrix} \meanA^{(1)} \\ \vdots \\ \meanA^{(\Ka)} \end{bmatrix} \quad \hspace{-0.2cm} \text{and} \quad
\CovA = \begin{bmatrix}
\CovA^{(1)}  	&\hspace{-0.2cm} \mathbf{0} 		&\hspace{-0.2cm} \cdots  & \mathbf{0} 		\\
\mathbf{0}  	&\hspace{-0.2cm} \CovA^{(2)} 		&\hspace{-0.2cm} \cdots  & \mathbf{0}   	\\
\vdots  	 	&\hspace{-0.2cm} \vdots  	      	&\hspace{-0.2cm} \ddots  & \vdots          	\\
\mathbf{0} 		&\hspace{-0.2cm} \mathbf{0} 		&\hspace{-0.2cm} \cdots  & \CovA^{(\Ka)}
\end{bmatrix},
\label{eq:alice_channel}
\end{equation}
where the diagonal structure of $\CovA$ follows from the assumption of independent fading across RRHs.
Now let us introduce the authentication test which is based on the generalized likelihood-ratio test (GLRT) often used in related work on PLA~\cite{Xiao2007,Xiao2008,Baracca2012}:

\begin{definition}[Authentication Hypothesis Test]
\label{def:glrt}
	 Bob makes an acceptance decision according to the following binary hypothesis test:
 \vspace{-2ex}\begin{equation}
	 d(\tilde{\h}_{m}) \mathop{\gtrless}_{\mathcal{H}_0}^{\mathcal{H}_1} T,
	 \label{eq:single_message_test}
\end{equation}
	where \textcolor{black}{$T$ is a descision threshold and} $d(\cdot)$ is a discriminant function given by
	
	 \vspace{-2ex}\begin{equation}
		d(\tilde{\h}_m) = 2\QF[\CovA^{-1}]{\tilde{\h}_m-\meanA}.
		\label{eq:discriminant}
	\end{equation}
\end{definition}

\textcolor{black}{Some comments are in order:}

\paragraph*{\textcolor{black}{Channel Requirements}}

\textcolor{black}{
The central prerequisite for the PLA scheme is the non-zero mean of the channel distribution $\CN(\mean_i^{(j)},\Cov_i^{(j)})$, which from a physical perspective is relevant when there are few dominating line-of-sight or reflective paths between transmitter and receiver (e.g., between sensor and access-points in a large open factory hall or between car and road-side access-points).
Such scenarios are common to assume within works on physical layer security~\cite{Abdelaziz2016,Abdelaziz2019,Yan2014,Forssell2019} and they can also be motivated by indoor channel measurements~\cite{MacLeod2005}.
The phased-array model of the channel mean in~\eqref{eq:phased_array} is not necessary for the function of the PLA scheme but rather a model we use to analyze the spatial properties of the scheme and to derive the optimal attack position.
To briefly address generalizations to non-LOS channels, we note that the non-zero mean distribution $\CN(\mean_i^{(j)},\Cov_i^{(j)})$ also could represent the predictive distribution of a channel predictor (e.g., a Kalman filter) used in conjunction with the PLA scheme.
}

\paragraph*{\textcolor{black}{Communication Overhead and Latency}}

\textcolor{black}{Intuitively, distributing multiple RRHs can improve the detection performance of the PLA scheme since impersonating the distance and AoA w.r.t. all antenna arrays becomes increasingly difficult.
Such benefits must however be weighed against the introduced system complexity.
}
In some cases, the CSI might only be locally available at the RRHs and not centrally available at Bob.
However, note that by defining $d^{(j)}(\tilde{\h}) = 2\QF[\{\Cov_{A}^{(j)}\}^{-1}]{\tilde{\h}^{(j)}-\mean_{A}^{(j)}}$ we can write the discriminant function in \eqref{eq:single_message_test} as $d(\tilde{\h}_m)=\sum_{j=1}^{\Ka} d^{(j)}(\tilde{\h}_m^{(j)})$, which holds due to the block diagonal structure of the covariance matrix $\CovA$.
\textcolor{black}{That is, in practice \eqref{eq:single_message_test} requires only the RRHs to communicate the soft decisions $d_i^{(j)}(\tilde{\h})$ (i.e., a single real value per authenticated message) which can be summed up at Bob, rather than communicating the CSI vector $\tilde{\h}$ (i.e., $\NRx$ complex values per authenticated message).}
\textcolor{black}{Note also that the distributed architecture might introduce latencies due to the links from the RRHs to Bob; however, this aspect is not considered in this work since the focus is on detection performance.}

\paragraph*{\textcolor{black}{Feature Learning}}
\label{par:discussion}

Note that prior to using the PLA scheme, Bob must initially learn the legitimate statistics $\meanA^{(j)},\CovA^{(j)}$.
This is an important problem as well as a common observation in related PLA literature, where it is often argued that the initial trust is established using cryptographic authentication whenever a new transmitter joins the network~\cite{Senigagliesi2020}.
Another solution in our scenario would be to use device position information to infer the corresponding channel statistics from the model~\eqref{eq:phased_array}.
Such an approach could also encompass device mobility by allowing the legitimate feature bank to be time-varying.
However, the details of such methods are considered outside the scope of this work, and we presuppose that Bob (or at least the RRHs) knows the time-invariant $\mean_i^{(j)},\Cov_i^{(j)}$ perfectly.

\vspace{-3ex}
\subsection{\textcolor{black}{Error Probabilities and Authentication Threshold}}
\label{sub:error_probabilities}

\textcolor{black}{
Two types of error events can occur in the binary authentication test in Definition~\ref{def:glrt}: false alarms and missed detections.
A false alarm is when a legitimate message is rejected, a missed detection is when an adversary message is accepted, and the probability of these events are defined as
$\pFA(T) = \prob(d(\tilde{\h}_{m})>T|\Hyp_0)$ and $\pMD(T) = \prob(d(\tilde{\h}_{m})<T|\Hyp_1)$, respectively.
It is easy to show that under the assumptions of this paper, $d(\tilde{\h}_{m})|\Hyp_0\sim\chi_{2\Ka\NRx}^2$, i.e., the discriminant function follows a central $\chi^2$ distribution whenever Alice is transmitting.
Hence, the false alarm probability can always be obtained in closed form according to $\pFA(T)=1-F_{\chi_{2\Ka\NRx}^2}(T)$.
The missed detection probability $\pMD(T)$ is generally not as straightforwardly tractable in the multiple RRH case since $d(\tilde{\h}_{m})|\Hyp_1$ is a weighted sum of non-central $\chi^2$ variables.
However, we have previously provided an efficient approximation in \cite{Forssell2019b} and we will provide solutions to this problem under optimal attack strategies throughout this paper.
}
\textcolor{black}{
The choice of authentication threshold $T$ is part of the system design.
Typically, one would start by determining a tolerable false-alarm rate $\pFA^{*}$ and compute the corresponding threshold
\begin{equation}
T^{*} = F_{\chi_{2\Ka\NRx}^2}^{-1}(1-\pFA^{*}),
\end{equation}
which can then be used to evaluate the security level $\pMD(T^{*})$.
}

\vspace{-3ex}

\subsection{PHY-Layer Attack Strategies}
\label{sub:adversarial_strategies}

Now we introduce the PHY layer attack strategies that are analyzed in Section~\ref{sec:opt_power_attack} and \ref{sec:opt_attack_pos}:

\begin{definition}[PHY Layer Attack 1: Power Manipulation]
\label{def:power_manipulation}
\label{def:PM}
Eve manipulates the transmit power and phase at her single-antenna transmitter by employing a complex scaling factor $\rho_E e^{j\varphi_E}$ such that the channel state observed at Bob becomes $\eta_E e^{j\psi_E}\hE$.
Eve can adopt either a fixed power manipulation strategy based on statistical CSI or a channel-realization dependent strategy based on perfect CSI at Eve.

\end{definition}

\begin{definition}[PHY Layer Attack 2: Attack Position]
\label{def:AP}
Eve chooses her spatial position with respect to the receive arrays to influence the statistics of her channel distribution.
The objective for Eve is to find the optimal position, i.e., the one that maximizes the missed detection probability with respect to the legitimate device position and the RRH deployment.

\end{definition}

These strategies can be launched by external entities (e.g., an attacker in close proximity to the system, using a stolen device or a software defined radio unit) or internal devices whose behavior has been hijacked by malicious code.
Obviously, these attacks can also be combined with MAC-layer attacks such as disassociation or Sybil attacks to maximize the attack impact.

\vspace{-3ex}

\subsection{Problem Formulation}
\label{sub:problem_formulation}

\textcolor{black}{
The choice of authentication threshold $T$ will influence the security (i.e., detection performance) and the system-level performance impacts (e.g., packet-drops and delays) of the PLA scheme due to the tradeoff between missed detections and false alarms.
Worst-case bounds on $\pMD(T)$ would allow a system designer to dimension the system parameters (i.e., number of arrays and antennas) and to calibrate the decision threshold in order to find an operation point with guaranteed system performance and security level.
The following sections are devoted to deriving such upper bounds under the PHY-layer attack strategies in Definition~\ref{def:PM} and~\ref{def:AP}.
Section~\ref{sec:opt_power_attack} provides the bound for the optimal power manipulation attack; i.e., we derive $\pMD^{(Opt. PMA)} = \max\limits_{\rho_E,\varphi_E}\pMD(T)$ for a given attack position.
Section~\ref{sec:opt_attack_pos} is devoted to the problem of finding the worst-case attack position; i.e., we maximize $\pMD^{(Opt. PMA)}$ over a set of allowed positions $\xiE\in\mathcal{R}$.
}

%

\vspace{-2ex}
\section{Power Manipulation Attack}		\label{sec:opt_power_attack}
In this section, we provide a worst-case missed detection performance analysis under the optimal power manipulation attack.
Recall that under a power manipulation strategy, Eve manipulates power and phase, modeled by complex scaling factor $\eta_{\text{E}} e^{j\psi_{\text{E}}}$, with the objective to maximize the success probability given by the probability of missed detection $\pMD(T)$.
First, we will derive the optimal strategy under the assumption that Eve has perfect knowledge of her instantaneous CSI and provide an approximation of the associated missed detection probability.
Next, we will introduce a strategy based on only statistical CSI knowledge and provide the missed detection probability under this assumption.

\vspace{-3ex}
\subsection{Optimal Attack Given Perfect CSI at Eve} 
\label{sub:power_manipulation_attack}

Here, we assume that Eve perfectly knows the channel states $\hE^{(j)}$ with respect to each RRH, prior to her impersonation attempt.
\textcolor{black}{We also assume that Eve has perfect knowledge of the legitimate feature statistics $\meanA$ and $\CovA$.}
\textcolor{black}{Such information could in practice leak to the attacker from the PLA feature-bank or be inferred based on knowledge of the positions, environment, and ray-tracing tools.}
This might be considered an unrealistically competent attacker; however, it is relevant since the missed detection probability under this assumption will serve as a worst-case upper bound for any power manipulation strategy.

Considering that the missed detection probability under the power manipulation attack is defined by $\pMD(T) = \prob(d(\eta_{\text{E}} e^{j\psi_{\text{E}}}\hE)<T)$, the optimal strategy will be to minimize the discriminant function $d(\eta_{\text{E}} e^{j\psi_{\text{E}}}\hE)$ given by \eqref{eq:discriminant}.
That strategy is provided in the following lemma:

\begin{lemma}[Optimal Power Manipulation Attack Given Perfect CSI]
	\label{lem:opt_power}
The power manipulation strategy that minimizes the discriminant function \eqref{eq:discriminant} is given by

 \vspace{-2ex}\begin{equation}
\eta_{\text{E}}^* = \frac{|\meanA^\dag\CovA^{-1}\hE|}{\hE^\dag\CovA^{-1}\hE}, \quad \psi_{\text{E}}^* = -\arg\{\meanA^\dag\CovA^{-1}\hE\},
\label{eq:optimal_attack_strategy}
\end{equation}
yielding the minimal achievable lower bound on the discriminant function $d(\hE) \geq d^{(\text{Opt. PMA})}$ where

 \vspace{-4ex}\begin{equation}
d^{(\text{Opt. PMA})} = 2\meanA^\dag\CovA^{-1}\meanA\left(1-\frac{|\meanA^\dag \CovA^{-1} \hE|^2}{\meanA^\dag\CovA^{-1}\meanA\hE^\dag\CovA^{-1}\hE}\right).
\label{eq:dmin}
\end{equation}
\end{lemma}

\begin{proof}
	See Appendix~\ref{app:optimal_power}.
\end{proof}

Strategy \eqref{eq:optimal_attack_strategy} allows us to formulate an upper bound for the detection performance in the following definition:

\begin{definition}[Detection Performance Under Optimal Power Manipulation Attack]	
\begin{equation}
	\begin{aligned}
	 \pMD \leq \pMD^{(\text{Opt. PMA})} \triangleq \prob(d^{(\text{Opt. PMA})}<T)
	= \prob\left( \frac{|\meanA^\dag \CovA^{-1} \hE|^2}{\meanA^\dag\CovA^{-1}\meanA\hE^\dag\CovA^{-1}\hE} > 1-\frac{T}{2\meanA^\dag\CovA^{-1}\meanA}\right).
	\label{eq:pMDwc1}
	\end{aligned}
	\end{equation}	
\end{definition}

To provide insight into the problem of evaluating \eqref{eq:pMDwc1}, we define $t = 1-\frac{T}{2\meanA^\dag\CovA^{-1}\meanA}$, $\z = \frac{\QA\meanA}{\|\QA\meanA\|}$, and $\bar{\h}_E=\QE^\dag\hE$, where $\Q_i$ is the Cholesky factorization of $\Cov_i^{-1}$ for $i=\{A,E\}$.
With some manipulation \eqref{eq:pMDwc1} can be re-written in terms of a quadratic form

 \vspace{-2ex}\begin{equation}
	\pMD^{(\text{Opt. PMA})}(T) =\prob(\bar{\h}_E^\dag\mathbf{A}(t)\bar{\h}_E>0)
	\label{eq:quadratic_form}
\end{equation}
where $\mathbf{A}(t) = \QE^{-1}\QA^\dag(\z\z^\dag-t\mathbf{I})\QA(\QE^{-1})^\dag$.

Note that the determinant $|\mathbf{A}(t)| = |\CovE||\CovA^{-1}||\z\z^\dag-t\mathbf{I}|=|\CovE||\CovA^{-1}|(-t)^{N-1}(1-t)$ which implies that, if $t>0$ and the total number of antennas $N = \Ka\NRx$ is even, $|\mathbf{A}(t)|<0$ and $\mathbf{A}(t)$ will have an odd number of negative eigenvalues. 
Hence, the probability $\pMD^{(\text{Opt. PMA})}(T)$ generally takes the form of the complementary CDF of an indefinite quadratic form in the complex Gaussian vector $\bar{\h}_E\sim\CN(\b,\mathbf{I})$ with $\b = \QE^\dag\meanE$. 
Closed-form expressions for such distributions are generally not tractable; however, several approximation methods exist in the literature.
In the following, we provide two efficient methods that can be used for evaluating the probability \eqref{eq:pMDwc1}.
First, we solve the problem in closed-form in Theorem~\ref{thm:single_array_mdp} for the single-array case ($\Ka=1$) by exploiting the particular structure of the matrix $\mathbf{A}(t)$.
Then we generalize the result to multiple arrays in Theorem~\ref{thm:sp_approximation}($\Ka>1$) based on a previously developed approximation for CDFs of indefinite quadratic forms \cite{Naffouri2016}.

\paragraph{Solution for $\Ka=1$}
\label{par:closed_form_wcpmd}
In the case of a single receive array, the worst-case missed detection probability can be evaluated in closed-form.
The reason is that under the assumption $\Ka=1$, we can analytically find the eigenvalues of the matrix $\mathbf{A}$, which allows us to write the statistic as a ratio of two $\chi^2$ random variables.
This ratio, by definition, follows a doubly non-central F-distribution for which closed-form distribution functions exist in the literature.
We formulate this result in the following theorem:

\begin{thm}[Single-Array Worst-Case Missed Detection Probability]
	\label{thm:single_array_mdp}
For a single receive array ($\Ka=1$), the worst-case missed detection probability can be obtained in closed form

 \vspace{-2ex}\begin{equation}
\begin{aligned}
\pMD^{(\text{Opt. PMA})}(T) = 1-F_{\text{DNCF}}\left(x;\nu_1,\nu_2,k_1,k_2\right)
\label{eq:pmd_wc}
\end{aligned}
\end{equation}
where 
$x=(\NRx-1)\left(1-\frac{2\meanA^\dag\CovA^{-1}\meanA}{T}\right)$, 
$\nu_1 =\frac{2|\meanA^\dag\CovA^{-1}\meanE|^2}{\alpha\meanA^\dag\CovA^{-1}\meanA}$, 
$\nu_2 = \frac{2}{\alpha}\left(\meanE^\dag\CovA^{-1}\meanE-\frac{|\meanA^\dag\CovA^{-1}\meanE|^2}{\meanA^\dag\CovA^{-1}\meanA}\right)$,
$k_1 = 2$, 
$k_2=2(\NRx-1)$, and

 \vspace{-4ex}\begin{equation}
F_{\text{DNCF}}\left(x;\nu_1,\nu_2,k_1,k_2\right) = e^{-\frac{\nu_1+\nu_2}{2}}\sum_{r=0}^{\infty}\sum_{s=0}^{\infty}\frac{\left(\frac{\nu_1}{2}\right)^r}{r!}\frac{\left(\frac{\nu_2}{2}\right)^s}{s!}I\left(\frac{k_1x}{k_2+k_1x};\frac{n_1}{2}+r,\frac{n_2}{2}+s\right)
\label{eq:DNCF}
\end{equation}
denotes the CDF of a doubly non-central F-distribution with non-centrality parameters $\nu_1$ and $\nu_2$, degrees of freedom $k_1$ and $k_2$, written in terms of the incomplete beta function $I(q;a,b) = \int_0^qt^{a-1}(1-t)^{b-1}dt$.

\end{thm}

\begin{proof}
We start from \eqref{eq:pMDwc1}, but instead define $\bar{\h}_E = \Q_A\h_E$ where $\Q_A$ again is the Cholesky factorization of $\Cov_A^{-1}$.
Using this, we can continue from \eqref{eq:pMDwc1} and write
 \vspace{-1ex}
\begin{equation}
\begin{aligned}
\pMD^{(\text{wc})}(T) 
&= \prob\left( \frac{|\z^\dag\bar{\h}_E|^2}{\|\bar{\h}_E\|^2}>t \right)
=  \prob\left( \bar{\h}_E^\dag\z^\dag\z\bar{\h}_E>t\bar{\h}_E^\dag\bar{\h}_E \right) 
= \prob(\bar{\h}_E^\dag(\z^\dag\z-t\mathbf{I})\bar{\h}_E>0).
\end{aligned}
\end{equation}
The matrix $\mathbf{A}_2 = \z^\dag\z-t\mathbf{I}$ is clearly Hermitian which means that we can write $\mathbf{A}_2 = \mathbf{U}^\dag\mathbf{D}\mathbf{U}$ where $\mathbf{D}$ is a real-valued diagonal matrix with the eigenvalues $d_i=-t$ for $i=1,\cdots,\NRx-1$ and $d_{\NRx}=1-t$ and $\mathbf{U}$ is an orthonormal matrix with the last column equal to $\z$. 
Now since $\Cov_E=\alpha\Cov_A$, we can let $\x\triangleq\mathbf{U}\bar{\h}_E = \mathbf{U}\Q_A\h_E \sim \CN(\mathbf{U}\Q_A\mean_E,\alpha \mathbf{I})$, $X_1\triangleq|x_{\NRx}|^2\sim \frac{2}{\alpha}\chi_{2}^{2}\left(\frac{2}{\alpha}|[\mathbf{U}\Q_A\mean_E]_{\NRx}|^2\right)$, and 
 \vspace{-1ex}
\begin{equation}
X_2 \triangleq \sum_{i=1}^{\NRx}|x_i|^2\sim\frac{2}{\alpha}\chi_{2(\NRx-1)}^{2}\left(  \frac{2}{\alpha}\sum_{i=1}^{\NRx}|[\mathbf{U}\Q_A\mean_E]_{i}|^2\right),
\end{equation}
which are independent due to the independence of the elements in $\x$.
Then we note that 
 \vspace{-2ex}
\begin{align}
\prob(\bar{\h}_E^\dag(\z^\dag\z&-t\mathbf{I})\bar{\h}_E>0) = \prob(\x^\dag\mathbf{D}\x>0) 	
= \prob\left((1-t)|x_{\NRx}|^2-t\sum_{i=1}^{\NRx}|x_i|^2>0\right)							\nonumber
\\
&= \prob\left( \frac{|x_{\NRx}|^2}{\sum_{i=1}^{\NRx}|x_i|^2}>\frac{t}{1-t}\right)  			
= \prob\left(\frac{\frac{X_1}{2}}{\frac{X_2}{2(\NRx-1)}}>(\NRx-1)\frac{t}{1-t}\right). \label{eq:pmd_wc2}
\end{align}
The lefthand-side ratio $Y \triangleq \frac{\frac{X_1}{2}}{\frac{X_2}{2(\NRx-1)}}$ in \eqref{eq:pmd_wc2} is therefore a ratio of normalized independent $\chi^2$ random variables which by definition follows a doubly non-central F-distribution. Hence, $\pMD^{(\text{wc})}(T) = \prob\left(Y>(\NRx-1)\frac{t}{1-t}\right)$ from which the result in \eqref{eq:pmd_wc} follows.
\end{proof}

\paragraph{Saddle-Point Approximation for $\Ka>1$}

In the multiple-array case, $\mathbf{A}$ will generally not possess the Hermitian property that was exploited in Theorem~\ref{thm:single_array_mdp}.
Therefore, we instead turn to integral approximation techniques.
Using the eigenvalue decomposition $\mathbf{A}=\U\D\U^\dag$ and a strategy similar to the one proposed in~\cite{Naffouri2016}, the probability \eqref{eq:pMDwc1} can be transformed into a one-dimensional integral, for any real-valued parameter $\beta>0$, stated in the following proposition

\begin{prop}[Alternative Formulation of Worst-Case Missed Detection Probability]
	 \begin{equation}
		\pMD^{(\text{Opt. PMA})} =-\frac{1}{2\pi}\int_{-\infty}^{\infty} \frac{e^{-c(\omega)}}{(\beta-j\omega)|\mathbf{I}+(\beta-j\omega)\D|}d\omega,
	\label{eq:int_rep}
	\end{equation}
	with the arbitrary real constant $\beta>0$, $\b = \QE^\dag\meanE$, and

	 \vspace{-2ex}\begin{equation}
		c(\omega) = \b^\dag\left(\mathbf{I} + \frac{1}{j\omega-\beta}\D^{-1}\right)^{-1}\b.
	\end{equation}
\end{prop}

\begin{proof}
	See Appendix~\ref{app:int_rep}.
\end{proof}

Although neither this integral is computable in closed form, it is easier to handle than the brute force $\Ka\NRx$-dimensional integral over the CSCG vector $\h$.
Here we use a saddle-point method to approximate the integral \eqref{eq:int_rep} in the following theorem:

\begin{thm}[Approximation of Worst-Case Missed Detection Probability for $\NRx\geq 2$]
\label{thm:sp_approximation}

The worst-case missed detection probability can be approximately evaluated as

 \vspace{-2ex}\begin{equation}
\pMD^{(\text{Opt. PMA})} \approx -\frac{1}{2\pi}e^{s(z_0)}e^{-j\angle s''(z_0)}\sqrt{\frac{2\pi}{|s''(z_0)|}},
\label{eq:sp_approximation}
\end{equation}
where
 \vspace{-2ex}\begin{equation}
	s(z) = - \b^\dag\left(\mathbf{I} + \frac{1}{z}\D^{-1}\right)^{-1}\b - \ln(z) - \ln(|\mathbf{I}+z\D|),
	\label{eq:s_z}
\end{equation}
$\b = \QE^\dag\meanE$, and $z_0$ is a stationary point such that $s'(z_0) = 0$.
\end{thm}

\begin{proof}
With a change to the complex variable $z=j\omega-\beta$, we can write \eqref{eq:int_rep} as

\begin{equation}
	\pMD^{(wc)} = -\frac{1}{j2\pi}\oint_{-\beta-j\infty}^{-\beta + j\infty} e^{-s(z)}dz
\end{equation}
with $s(z)$ defined according to \eqref{eq:s_z}. The saddle point method uses the approximation $s(z)\approx s(z_0) + \frac{1}{2}s''(z_0)(z-z_0)^2$ to write

\begin{equation}
\begin{aligned}
\pMD^{(wc)} &\approx -\frac{1}{j2\pi}\oint_{-\beta-j\infty}^{-\beta + j\infty} e^{-(s(z_0) + \frac{1}{2}s''(z_0)(z-z_0)^2)}dz \\
&= -\frac{1}{j2\pi}e^{-s(z_0)}\oint_{-\beta-j\infty}^{-\beta + j\infty} e^{- \frac{1}{2}s''(z_0)(z-z_0)^2}dz 
= -\frac{1}{j2\pi}e^{-s(z_0)} e^{j\phi} \sqrt{\frac{2\pi}{|s''(z_0)|}}
\end{aligned}
\end{equation}
with $\phi = \frac{\pi-\angle s''(z_0)}{2}$. Finally, we note that $e^{j\phi} = je^{-j\angle s''(z_0)}$ from which \eqref{eq:sp_approximation} follows.

\end{proof}

\vspace{-3ex}
\subsection{Fixed Power Manipulation Strategy (Statistical CSI at Eve)} 

Suppose now that Eve can only choose a fixed strategy for $\eta_{\text{E}} e^{j\psi_{\text{E}}}$, i.e., one that does not depend on the instantaneous CSI $\hE$.
For example, Eve can choose a strategy based on knowledge of $\meanA$ and $\meanE$, which in practice could be obtained by using ray-tracing tools.
For any fixed strategy, we can clearly formulate the missed detection probability as

 \vspace{-2ex}\begin{equation}
	\pMD^{(\text{Fixed PMA})}(T) = \prob(\QF[\CovA^{-1}]{\eta_{\text{E}} e^{j\psi_{\text{E}}}\hE-\meanA}<T/2)
	\label{eq:fixed_pow_mdp}
\end{equation}
Note that we have $\eta_{\text{E}} e^{j\psi_{\text{E}}}\hE-\meanA\sim\CN(\mean,\Cov)$ with $\mean = \eta_{\text{E}}e^{j\psi_{\text{E}}}\meanE-\meanA$ and $\Cov = \eta_{\text{E}}^2\CovE$, so the probability \eqref{eq:fixed_pow_mdp} again takes the form of a CDF of a complex Gaussian quadratic form.
Hence, we can calculate \eqref{eq:fixed_pow_mdp} by using the following corollary to Theorem~\ref{thm:sp_approximation}:

\begin{cor}
	\label{cor:fixed_pma}
	For the fixed power manipulation strategy, we get $\pMD^{(\text{Fixed PMA})}(T)$ by replacing $\bar{\h}_E$ with $\eta_{\text{E}} e^{j\psi_{\text{E}}}\hE-\meanA$ in \eqref{eq:quadratic_form} and apply the saddle-point approximation as described in Theorem~\ref{thm:sp_approximation}.
\end{cor}

Note that the approach in Corollary~\ref{cor:fixed_pma} also allows us to evaluate the missed detection probability without power manipulation attack by letting $\eta_{\text{E}}=1$ and $\psi_{\text{E}}=0$.

Finally, in the following definition we provide a special case of fixed strategy when Eve has only statistical CSI knowledge:

\begin{definition}[Power Manipulation Attack Based On Statistical CSI]
\begin{equation}
\eta_{\text{E}}^{\text{(stat)}} = \frac{|\meanA^\dag\CovA^{-1}\meanE|}{\meanE^\dag\CovA^{-1}\meanE}, \quad \psi_{\text{E}}^{\text{(stat)}} = -\arg\{\meanA^\dag\CovA^{-1}\meanE\},
\label{eq:fixed_pow_strat}
\end{equation}	
\end{definition}

The motivation behind strategy \eqref{eq:fixed_pow_strat} is to use the strategy derived in Lemma~\ref{lem:opt_power} but assume the strong LoS approximation $\hE\approx\meanE$.

%

\vspace{-2ex}
\section{Optimal Attack Position}		\label{sec:opt_attack_pos}

In this section, we study the problem of finding the optimal attack position with respect to a given deployment.
We will denote Alice's and Eve's spatial positions by $\xiA$ and $\xiE$, respectively, and assume that Eve knows the deployment and position of the legitimate device.
Obviously, a straightforward solution for optimizing $\pMD^{(\text{Opt. PMA})}(\xiE)$ is to pick $\xiA=\xiE$ which will result in $\pMD^{(\text{Opt. PMA})}=1-\pFA$.
However, a basic underlying assumption is that the attacker is significantly separated from the legitimate device since the authentication method in-itself is predicated on this spatial separation in order to work.
Therefore, here we rather seek locally optimal attacker positions that are outside the immediate neighborhood of the legitimate device.
We start with defining and characterizing the objective function and identify properties of local optima that we exploit in our heuristic search algorithm. 
The algorithm is then presented in Section~\ref{sub:heuristic_solution_for_general_rice_fading}.

\vspace{-3ex}
\subsection{General Optimization Problem}
\label{sub:optimization_problem}

The optimal attack position is equivalent to the one maximizing $\pMD^{(\text{Opt. PMA})}(\xiE) = \prob(d^{(\text{Opt. PMA})}(\hE)<T)$, that is, the worst-case position given that Eve is using the optimal power manipulation attack.
Straightforwardly, from rearranging \eqref{eq:pMDwc1}, this problem can be rewritten in a convenient form, as stated in the following definition:

\begin{definition}[Optimal Position Attack]
	\label{def:opt_pos_attack}
	We define the region of allowed attack positions as $\mathcal{R}$ and let $\meanE(\xiE),\CovE(\xiE)$ denote the channel statistics induced by the attack position $\xiE\in\mathcal{R}$.
The optimal attack position is given as the solution to
	
	\begin{equation}
		\xiE^* = \argmax_{\xiE\in\mathcal{R}} \prob(F_{\text{obj}}(\hE)>T^*),
		\label{eq:opt_problem}
	\end{equation}
where
	\begin{equation}
	F_{\text{obj}}(\hE) = \frac{|\meanA^\dag \CovA^{-1} \hE|^2}{\hE^\dag\CovA^{-1}\hE}
	\label{eq:fobj}
	\end{equation}
is an objective function and $T^*=\meanA^\dag\CovA^{-1}\meanA-T/2$ is a constant threshold.	
\end{definition}

Direct optimization of \eqref{eq:opt_problem} results in a very complicated optimization problem due to the somewhat complicated distribution of \eqref{eq:fobj}.

\vspace{-3ex}
\subsection{Characterization of Objective Function Under Strong LoS Assumption} 

First, let us consider the case of strong LoS conditions, i.e., when $\KRice$ is large and $\hE\approx\meanE(\xiE)$.
In such a setting, the objective function in \eqref{eq:fobj} is approximately $F_{\text{obj}}(\meanE(\xiE))$.
Under these assumptions, our approach is to expand \eqref{eq:fobj} to provide an understanding of how the missed detection probability depends on the attack position.
First, we introduce some notation related to the positions of Alice and Eve with respect to the RRHs that will prove useful:
We define the distance ratios $r_j = \frac{d_{\text{A}}^{(j)}}{d_{\text{E}}^{(j)}}$, the phase offsets $\varphi_{\text{E}}^{(j)}=\frac{2\pi d_{\text{E}}^{(j)}}{\lambda_c}$ and $\varphi_{\text{A}}^{(j)}=\frac{2\pi d_{\text{A}}^{(j)}}{\lambda_c}$, phase differences $\Delta\varphi_j = \varphi_{\text{E}}^{(j)}-\varphi_{\text{A}}^{(j)}$, and angular-sine differences $\DOk= \Omega_{\text{E}}^{(j)}-\Omega_{\text{A}}^{(j)}$.
Furthermore, we define the per-array inner products of the angular responses as

\begin{equation}
	S_{ik}^{(j)} = \e(\Omega_i^{(j)})^\dag\A^{-1}\e(\Omega_k^{(j)}),
	\label{eq:ang_inn_prod}
\end{equation}
for $i,k\in\{\text{A},\text{E}\}$.

For certain correlation matrices, we can additionally expand the inner products $\SEAk$ according to the following lemma: 

\begin{lemma}
	\label{lem:SEAk}
For any correlation matrix with inverse in the form $\A^{-1} = \mathbf{T} + \mathbf{M}$, where $\mathbf{T}$ is a symmetric Toeplitz matrix defined by the first column $[t_0,...,t_{N-1}]^T$ and $\mathbf{M}$ is a diagonal matrix with $[\mathbf{M}]_{k,l} = m_0$ for $k=l=2,...,N-1$ (i.e., all diagonal elements equal except  the first and last one being zero), we have

\begin{equation}
\SEAk = e^{j2\pi\frac{\NRx-1}{2}\Delta_r \DOk}g(\Omega_E^{(j)}),
\label{eq:SEAk}
\end{equation}
where $g(\Omega_E^{(j)})$ is a real-valued function.
\end{lemma}

\begin{proof}
	The sum along the main diagonal in $\mathbf{T}$ will take the form $\sum_{n=0}^{\NRx-1}t_0 e^{j2\pi\Delta_r \DOk n}$, which possesses the property in \eqref{eq:SEAk} according to exponential sum formulas. 
	This property extends to all the remaining diagonals in $\mathbf{T}$ and $\mathbf{M}$ but the details are left out due to space limitations.
\end{proof}

We provide the expanded representation of $F_{\text{obj}}(\meanE(\xiE))$ in the following lemma:

\begin{lemma}[Expanded Objective Function]
	\label{lem:exp_obj_fun}
\begin{equation}
	F_{\text{obj}}(\meanE(\xiE)) = \KRice \left|\frac
	{\sum_{j=1}^{\Ka}\sqrt{\bar{r}_j}|g(\Omega_E^{(j)})|e^{j\phi_0^{(j)}}}
	{\sqrt{\sum_{j=1}^{\Ka}\bar{r}_j\SEEk}} \right|^2
	\label{eq:exp_fobj}
\end{equation}
with $\phi_0^{(j)} = \Delta\varphi_j+2\pi\frac{\NRx-1}{2}\Delta_r \DOk + \frac{\pi}{2}[\text{sign}\{g(\Omega_E^{(j)}\}-1]$
and $\bar{r}_j = \frac{r_j^{\beta}}{\sum_{l}^{\Ka}r_l^{\beta}}$ where $\beta$ is the path-loss exponent.
\end{lemma}

\begin{proof}
	We obtain \eqref{eq:exp_fobj} by expanding \eqref{eq:fobj} using the block diagonal structure of $\CovA$, the definitions of $\meanE$ and $\meanA$ according to \eqref{eq:phased_array}, and the result in Lemma~\ref{lem:SEAk}.
\end{proof}

By inspecting \eqref{eq:exp_fobj}, we can make two observations that we will exploit in Section~\ref{sub:heuristic_solution_for_general_rice_fading}:
\begin{enumerate}
	\item Small-scale optimization of $F_{\text{obj}}(\meanE(\xiE))$ depends on the complex coefficients $e^{j\phi_0^{(j)}}$ related to the phase-relation of transmissions received at each RRH.
	\item Large-scale optimization depends on the angular responses $|g(\Omega_E^{(j)})|$ and the normalized distance ratios $\bar{r}_j$.
\end{enumerate}

\vspace{-3ex}
\subsection{Impact of Fading Correlation} 
\label{sub:fading_correlation}
In addition to the general result in Lemma~\ref{lem:SEAk}, we provide the angular response $g(\DOk)$ in closed form under two special cases, summarized in the two following lemmas:
\begin{lemma}[Uncorrelated Antennas]
	\label{lem:SEAk2}
	For uncorrelated antennas ($\A = \mathbf{I}$), we get	
	\begin{equation}
		g(\DOk) = \frac{\sin(\pi\Delta_r\NRx\DOk)}{\NRx\sin(\pi\Delta_r\DOk)}.
	\end{equation}
\end{lemma}

\begin{proof}
This follows from the conceptual proof of Lemma~\ref{lem:SEAk} with $t_0=1$ and $t_i=0$ for $i>0$.
\end{proof}

\begin{lemma}
	\label{lem:SEAk3}
For the exponential correlation matrix $\A_{k,l} = \rho^{-|k-l|}$, we have

\begin{equation}
\begin{aligned}
g(\DOk) = \frac{1}
					{(1-\rho^2)\sin(\pi\Delta_r\DOk))} \\
\times \big[ \sin(\pi\Delta_r\NRx\DOk)+
\rho^2\sin(\pi\Delta_r(\NRx-2)\DOk)
\\-2\rho\cos(\pi\Delta_r(\Omega_{E,k}+\Omega_{A,k}))\sin(\pi\Delta_r(\NRx-1)\DOk \big].
\end{aligned}
\end{equation}
\end{lemma}

\begin{proof}
The inverse of the exponential correlation matrix is a Toeplitz matrix with $t_0 = \frac{1}{1-\rho^2}$, $t_1 = \frac{\rho^2}{1-\rho^2}$, $t_2 = \frac{-2\rho}{1-\rho^2}$, and $t_i=0$ for $i>2$.
The rest follows similarly to the proof of Lemma~\ref{lem:SEAk}.
\end{proof}

\vspace{-3ex}
\subsection{Characterization of Locally Optimal Attack Positions for $\A = \mathbf{I}$ and $\Ka=2$} 
\label{par:example_ka_2_with_uncorrelated_antennas}
To simplify the analysis, we assume a deployment of two RRHs and uncorrelated antenna fading 
(i.e., $\A = \mathbf{I}$).
In the case of $\A = \mathbf{I}$, it is easy to find that $\SEEk=1$ and, thus, we have $\sum_{k=1}^{\Ka}\bar{r}_k\SEEk=1$.
Moreover, with the assumption of $\Ka=2$, we can write the expanded objective function as
\begin{equation}
	F_{\text{obj}}(\meanE(\xiE))=\left|\sqrt{\bar{r}_1}|g(\Delta\Omega_1)|e^{j\phi_0^{(1)}}+\sqrt{\bar{r}_2}|g(\Delta\Omega_2)|e^{j\phi_0^{(2)}}\right|^2.
		\label{eq:exp_fobj_Ka2}
\end{equation}
The small-scale local optima allow us to reduce the optimization search to a set of spatial sampling points for which we have a specific phase relation as specified in the following lemma:

\begin{lemma}[Small-Scale Spatial Sampling $\Ka=2$]
	The small-scale local optima of $F_{\text{obj}}(\meanE(\xiE))$ are found at points where $e^{j\phi_0^{(1)}}=e^{j\phi_0^{(2)}}$. At such points we have
	\begin{equation}
		F_{\text{obj}}(\meanE(\xiE)) =(\sqrt{\bar{r}_1}|g(\Delta\Omega_1)|+\sqrt{\bar{r}_2}|g(\Delta\Omega_2)|)^2.
			\label{eq:large_scale_fobj}
	\end{equation}
	\label{lem:small_scale}
\end{lemma}
\begin{proof}
Clearly, \eqref{eq:exp_fobj_Ka2} is maximized when $\arg(\sqrt{\bar{r}_k}|g(\Omega_E^{(j)})|e^{j\phi_0^{(j)}}) = \phi_0^{(j)}=\phi_0$ for $j=1,2$.
\end{proof}

\begin{remark}
Lemma~\ref{lem:small_scale} generalizes to $\Ka>2$ by considering points where $e^{j\phi_0^{(1)}}=e^{j\phi_0^{(2)}}=\cdots =e^{j\phi_0^{(\Ka)}}$. 
However, note that the existence of points with optimal phase alignment with respect to more than two arrays at a time is not guaranteed and depends on the RRH deployment and Alice's position. 
\end{remark}

Now let us restrict the angular sine differences to the set $\Delta\Omega\in\mathcal{A}$, where $\mathcal{A}$ is a set of local optima of the angular response $|g(\Delta\Omega)|$.

\begin{remark}
	Note that for RRH $j$, we have $\Delta\Omega = \sin(\Phi_E^{(j)})-\sin(\Phi_A^{(j)})$ which implies that each local optima $\Delta\Omega\in\mathcal{A}$ is associated with two attack angles $\Phi_E^{(j,+)}=\sin^{-1}(\Delta\Omega+\sin(\Phi_A^{(j)}))$ and $\Phi_E^{(j,-)}=\pi-\Phi_E^{(j,+)}$.
\end{remark}

Now we can characterize the large-scale local optima in the following theorem:

\begin{thm}[Large-Scale Local Optima for $\Ka=2$ arrays]
	\label{thm:optimal_point}
	At far-field points where $\bar{r}_k$ remain approximately constant in the local neighborhood, large-scale local optima of $F_{\text{obj}}(\meanE(\xiE))$ are found at the intersection points of the set lines with AoAs associated with angular sines $\Delta\Omega\in\mathcal{A}$.
\end{thm}
\begin{figure*}[h]
\centering
\psfrag{A}[][]{\small Alice Position}
\psfrag{B}[][]{\small RRH1}
\psfrag{C}[][]{\small RRH2}
\includegraphics[width=0.7\textwidth]{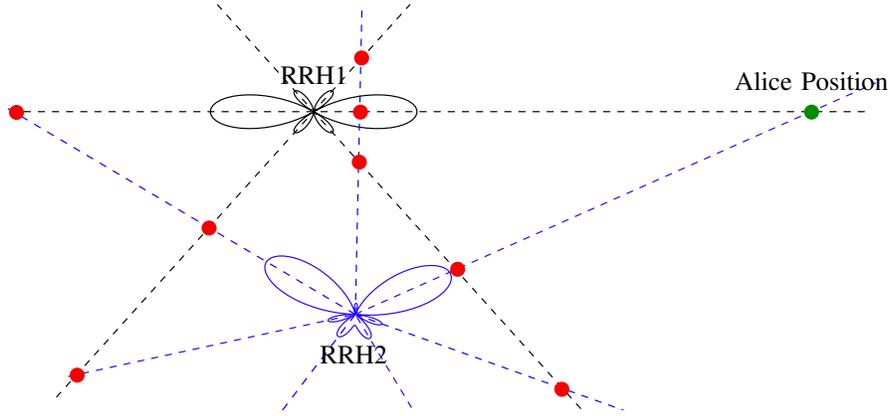}
\caption{Illustration of candidate points for optimal attacker position with $\Ka=2$ RRHs. Dashed lines indicate the rays with AoA $\Phi_{E,l}^{*}$.}
\label{fig:optimal_points}
\end{figure*}

\begin{proof}
Lets choose two particular angular sines $\Delta\Omega_1,\Delta\Omega_2\in\mathcal{A}$.
Let $\xiE^{*}$ denote the intersection point between the lines\footnote{Generally, such an intersection point might not exist; however, for the proof of this theorem we assume that $\Delta\Omega_1$ and $\Delta\Omega_2$ are chosen such that it does.} with AoAs $\Phi_E^{(1,+)}$ and $\Phi_E^{(2,+)}$.
Now from the definition of $\mathcal{A}$, we know that $|g(\Delta\Omega_k+\epsilon_k)|<|g(\Delta\Omega_k)|$ for $k=1,2$ and $\epsilon_k$ sufficiently small so that we stay in the neighborhood of the local optima of $g(\cdot)$.
If we deviate from $\xiE^{*}$ to a point $\xiE'$ with any angle offsets $\epsilon_1$ and $\epsilon_2$, where the deviation is small such that $\bar{r}_k$ is approximately constant, then

\begin{align}
	F_{\text{obj}}(\meanE(\xiE'))=  \KRice \left(\sqrt{\bar{r}_1}|g(\Delta\Omega_1+\epsilon_1)| + \sqrt{\bar{r}_2}|g(\Delta\Omega_2+\epsilon_2)| \right)^2 \\
			< \KRice \left(\sqrt{\bar{r}_1}|g(\Delta\Omega_1)| + \sqrt{\bar{r}_2}|g(\Delta\Omega_2)| \right)^2 =F_{\text{obj}}(\meanE(\xiE^{*})),
\end{align}
which shows that $\xiE^{*}$ is a large-scale local optima of $F_{\text{obj}}(\meanE(\xiE))$.
\end{proof}	
Fig.~\ref{fig:optimal_points} provides an illustration of the intersection points considered in Theorem~\ref{thm:optimal_point}.

The analysis thus far provides us with characterizations of small- and large-scale locally optimal points under certain assumptions.
In the final part of this section, we will exploit these results to develop a heuristic truncated search algorithm that can be used to find the optimal attack position in the general case efficiently.

\subsection{Heuristic Search Method for General Deployments and Rice Fading} 
\label{sub:heuristic_solution_for_general_rice_fading}

Our proposed search method can be summarized as follows: (i) In accordance with Theorem~\ref{thm:optimal_point}, reduce the search to points where AoAs are within the main lobe of a RRH or intersections of $1^{\text{st}}$ order side-lobes.
(ii) Based on Lemma~\ref{lem:small_scale}, use the function $F_{\text{small-scale}}(\xiE)$ defined below to find small-scale locally optimal points. (iii) Compute the missed detection probability $\pMD^{(\text{Opt. Position})}$ for the truncated set of small-scale local optima from step (ii).

The search algorithm is based on the following definitions:
We let 

	\begin{equation}
		F_{\text{small-scale}}(\xiE) = \left|\sum_{k=1}^{\Ka}e^{j\phi_0^{(j)}(\xiE)}\right|
		\label{eq:small_scale_obj}
	\end{equation}
be a small-scale optimization function for finding small-scale locally optimal points and let $B(\xiE,\epsilon)$ define the set of points within distance $\epsilon$ from attack position $\xiE$.
For RRH $j$, the main lobe AoAs are $\Phi_{\text{main}}^{(j)}=\{\Phi_A^{(k)},\pi-\Phi_A^{(k)}\}$ and we let $\Phi_{1^{\text{st}}}^{(j)}$ denote the first side lobe AoA (local maxima) of the angular response.
Based on this, the sets of searched AoA are defined as $\mathcal{A}_{\text{main}}^{(j)}=[\Phi_{\text{main}}^{(j)}-\delta_{-},\Phi_{\text{main}}^{(j)}+\delta_{+}]$ where $\delta_{+/-}$ is chosen such that 
$\left|g\left(\sin(\Phi_{\text{main}}^{(j)})-\sin(\Phi_A^{(k)})\right)\right|=g_0 \left|g\left(\sin(\Phi_{\text{main}}^{(j)}\pm\delta)-\sin(\Phi_A^{(k)})\right)\right|$, for a constant $g_0$. 
The AoA search set for the first side lobe $\mathcal{A}_{1^{\text{st}}}^{(j)}$ is similarly defined.
Based on these definitions, Algorithm~\ref{alg:truncated_search} describes the steps of the search method in mathematical detail.
In related work, sometimes the main lobe beam width is defined as $2/L_r$ where $L_r = \lambda_c\Delta_r\NRx$ represents the array length.
This definition can also be used in our problem, but note that the parametric choice based on $g_0$ is more general as it allows us to tune the beam width considered in the search.

\begin{algorithm}
\caption{Truncated Search: Critical Attack Positions}
\begin{algorithmic}[1]
%
%

\Procedure{FindOptimalAttackPosition}{$\mathcal{R}$} \Comment{Search allowed region $\mathcal{R}$}

\For {$j=1,\cdots,\Ka$}
	
	\State $\mathcal{P}\gets\{\xiE\in\mathcal{R};\Phi_E^{(j)}\in\mathcal{A}_{\text{main}}^{(j)}\}$
	\Comment{Mainlobes}
	
	\For {$k=1,\cdots,\Ka$ and $k\neq j$}
		\State $\mathcal{P}\gets \mathcal{P} \cup
		\{\xiE\in\mathcal{R};\Phi_E^{(j)}\in\mathcal{A}_{1^{\text{st}}}^{(j)}\land\Phi_E^{(k)}\in\mathcal{A}_{1^{\text{st}}}^{(k)}\}$ 	\Comment{Intersect. of sidelobes}
	\EndFor
	
	\State $\mathcal{P}_{\text{Critical}}(j) \gets \{\xiE\in\mathcal{P};F_{\text{small-scale}}(\xiE)\geq F_{\text{small-scale}}(\xi)\forall\xi\in B(\xiE,\epsilon) \}$\\\Comment{Restrict to small-scale local optima}
	
\EndFor

\State  $\pMD^{(\text{Opt. Position})}= \max_{\xiE\in\cup_{j=1}^{\Ka}\mathcal{P}_{\text{Critical}}(j)} \pMD^{(\text{Opt. PMA})}(\xiE)$

\EndProcedure
\end{algorithmic}
\label{alg:truncated_search}
\vspace{-4ex}
\end{algorithm}

At a first glance, it may appear as an arbitrary choice to restrict the search to the main lobes and intersections of $1^{\text{st}}$ order side lobes.
However, the structure of the maxima of the angular responses in Lemma~\ref{lem:SEAk2} and~\ref{lem:SEAk3} induces a decreasing hierarchy of side lobe maxima. 
Unfortunately, an analytical characterization of the objective function with respect to the impact of the local maxima of $g(\cdot)$ is difficult, since distances change with intersection points as well. 
However, our numerical results presented in the next section confirm that it is sufficient to restrict the search to intersections of first-order side lobes since local optima of higher order side lobes are inferior.

\vspace{-2ex}
\section{Numerical Results}		\label{sec:num_res}
\begin{figure}
\centering
\includegraphics[width=0.5\columnwidth]{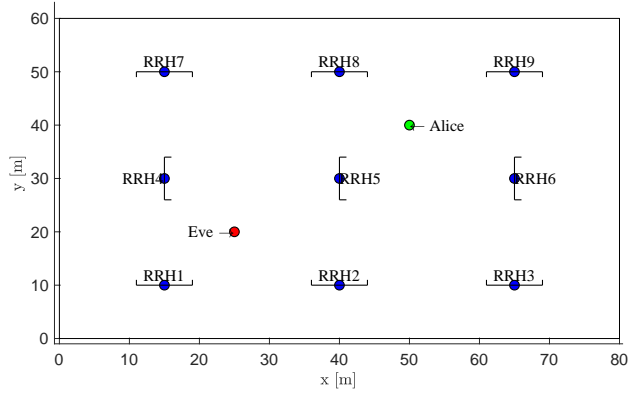}
\vspace{-2ex}
\caption{The 80 m$\times$60 m network deployment area used for numerical evaluations: 9 fixed RRH locations A1-A9, legitimate transmitter device Alice, and the attacker Eve.}
\vspace{-3ex}
\label{fig:deployment}
\end{figure}

In this section, we numerically study the detection performance under the considered attack strategies for different distributed network topologies.
We consider a system deployed in a 80 m$\times$60 m area, as depicted in Fig.~\ref{fig:deployment}.
In the area there are 9 potential RRH locations RRH1-RRH9 where antenna arrays of varying sizes can be placed, a legitimate device Alice, and the attacker Eve.
For all the subsequent results, we assume that the normalized antenna separation $\Delta_r=0.5$, the path-loss exponent $\beta=2$, and that the system operates at $f_c=2.4$ GHz center frequency.
Moreover, for the following results the authentication threshold $T$ is fixed for $\pFA = 10^{-2}$ unless stated otherwise.
\textcolor{black}{All results are evaluated based on channel distributions computed from the relative positions in the two-dimensional area of Fig.~\ref{fig:deployment} and the phased-array model \eqref{eq:phased_array}.
}

\vspace{-3ex}
\subsection{Validation of Saddle-Point Approximation} 
\label{sub:validation_of_saddle_point_approximation}

\begin{figure*}
\centering
\begin{subfigure}{0.5\textwidth}\psfrag{X}[][]{\small False alarm prob. $\pFA$}
\psfrag{Y}[][][1][-180]{\small Missed detection prob. $\pMD$}
\includegraphics[width=\columnwidth]{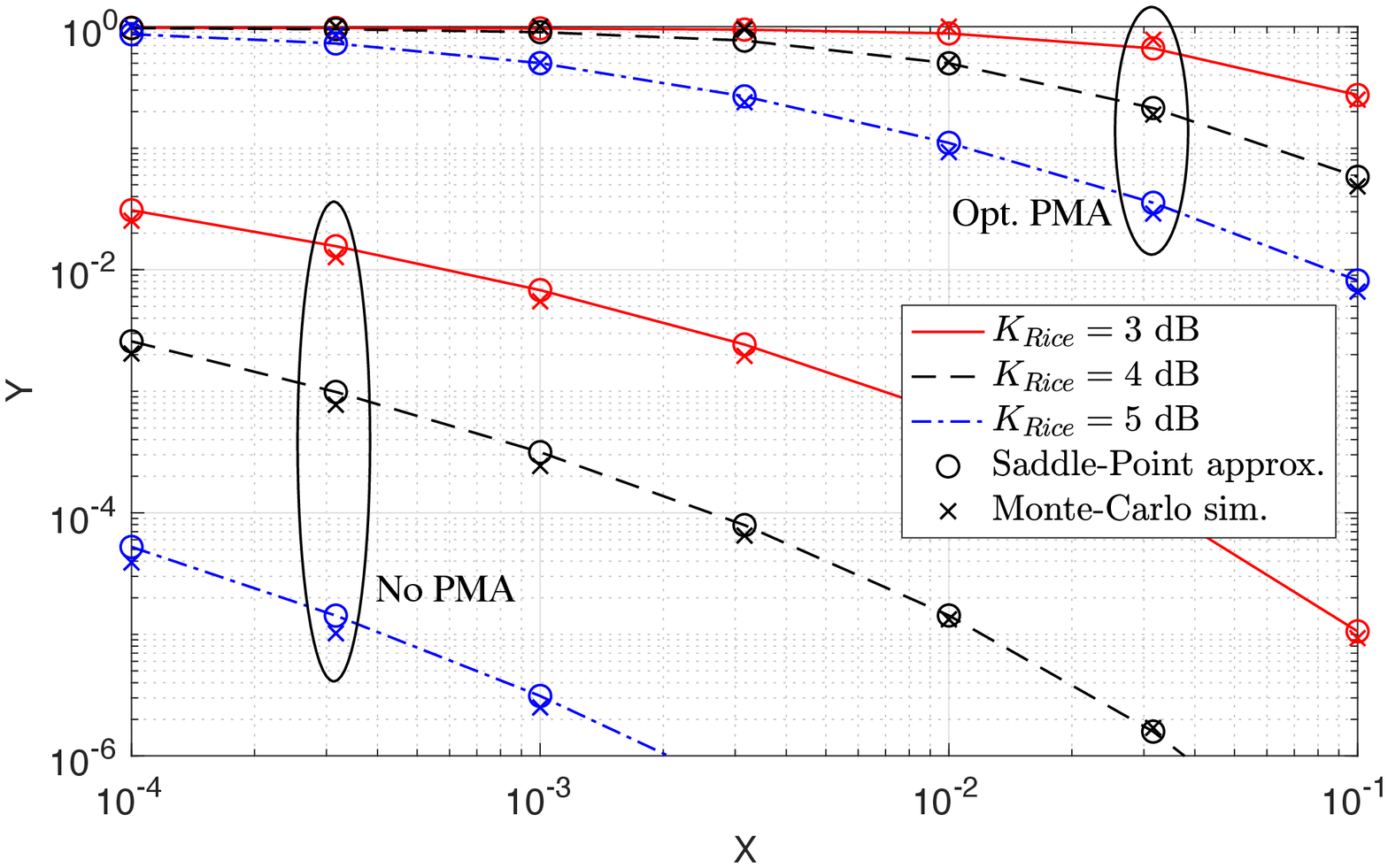}
\caption{}
\label{fig:ROC}
\end{subfigure}%
\begin{subfigure}{0.5\textwidth}
\psfrag{X}[][]{\small Fading correlation coeff. $\rho$}
\psfrag{Y}[][][1][-180]{\small Missed detection prob. $\pMD$}
\includegraphics[width=\columnwidth]{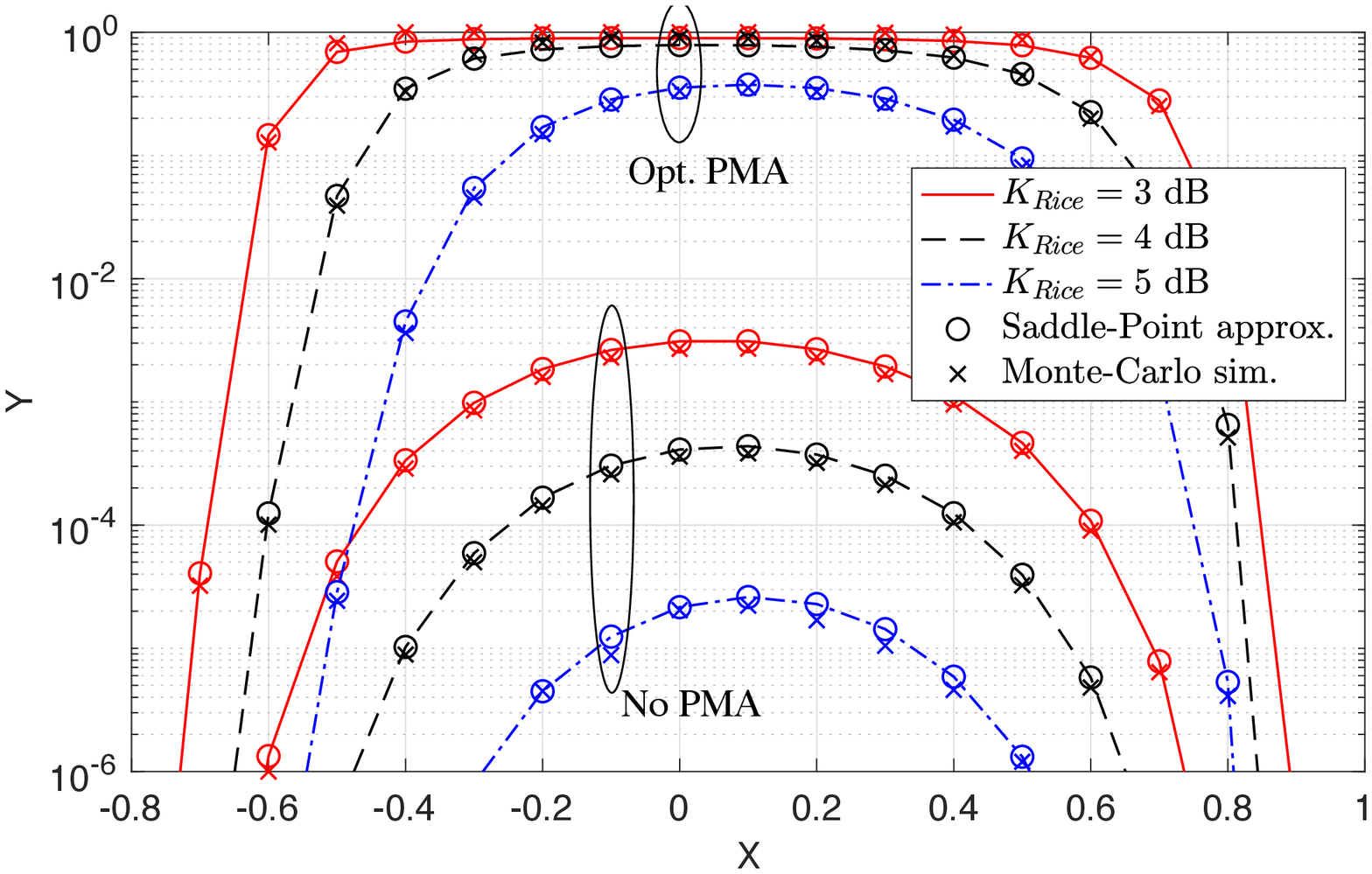}
\caption{}
\label{fig:CORR}
\end{subfigure} %
\vspace{-2ex}
\caption{Saddle-point approximation of $\pMD^{\text{(Opt. PMA)}}$ compared to Monte-Carlo simulations for a $\Ka=3$ RRH deployment for: (a) different false alarm probabilities and (b) varying correlation coefficient $\rho$.}
\vspace{-3ex}
\label{fig:validation_1}
\end{figure*}

\textcolor{black}{
In Fig.~\ref{fig:validation_1}, we validate the approximation of the missed detection probability (MDP) under the power manipulation attack by comparing it to Monte-Carlo simulations.}
We plot the missed detection probability both under optimal power manipulation, i.e., $\pMD^{\text{(Opt. PMA)}}$ and without power manipulation.
The curves represent the evaluation of the saddle-point approximation in Theorem~\ref{thm:sp_approximation}, and the Monte-Carlo simulation results, indicated by the cross markers, are computed based on $10^{7}$ channel realizations sampled from CSCG distributions. 
The results are based on a deployment of $\Ka=3$ RRHs (RRH locations 1, 3, and 8) with $\NRx=2$ antennas each.
Alice and Eve are located at $\xi_A=(\text{65 m},\text{30 m})$ and $\xi_E=(\text{26 m},\text{49 m})$, respectively.
\textcolor{black}{
We note from Fig.~\ref{fig:validation_1} that the proposed saddle-point approximation provides accurate results for the illustrated parameter ranges.}
Small approximation errors can be observed which generally seem to upper bound the values obtained from the simulations.
\textcolor{black}{
In addition, we have validated the approximation for varying position and number of antennas $\NRx$ (for increasing $\NRx$, the detection performance improves as expected); however these results are omitted due to space constraints.}
Fig.~\ref{fig:ROC} shows the missed detection probability for varying false alarm probabilities (i.e., this is the receiver operating characteristic curve for varying choices of threshold $T$).
Observe that Eve gains significantly (i.e., several orders of magnitude) in success probability by using the optimal attack strategy.
Fig.~\ref{fig:CORR} illustrates the missed detection probability for varying values of the fading correlation coefficient $\rho$.
We generally observe a decreasing missed detection probability as the magnitude $|\rho|$ increases.
Note that the fading correlation coefficient $\rho$ represents the correlation in the channel fading; i.e., $\rho=0$ represents independent fading  across antennas and $\rho=1$ represents full correlation.

%

\vspace{-3ex}
\subsection{Impacts of Power Manipulation Attack} 

\begin{figure*}
\centering
\begin{subfigure}{0.5\textwidth}
\psfrag{XA}[][]{\small Distance to RRH4 $d_E^{(1)}$ [m]}
\psfrag{YA}[][]{\small Missed detection prob. $\pMD$}
\includegraphics[width=\columnwidth]{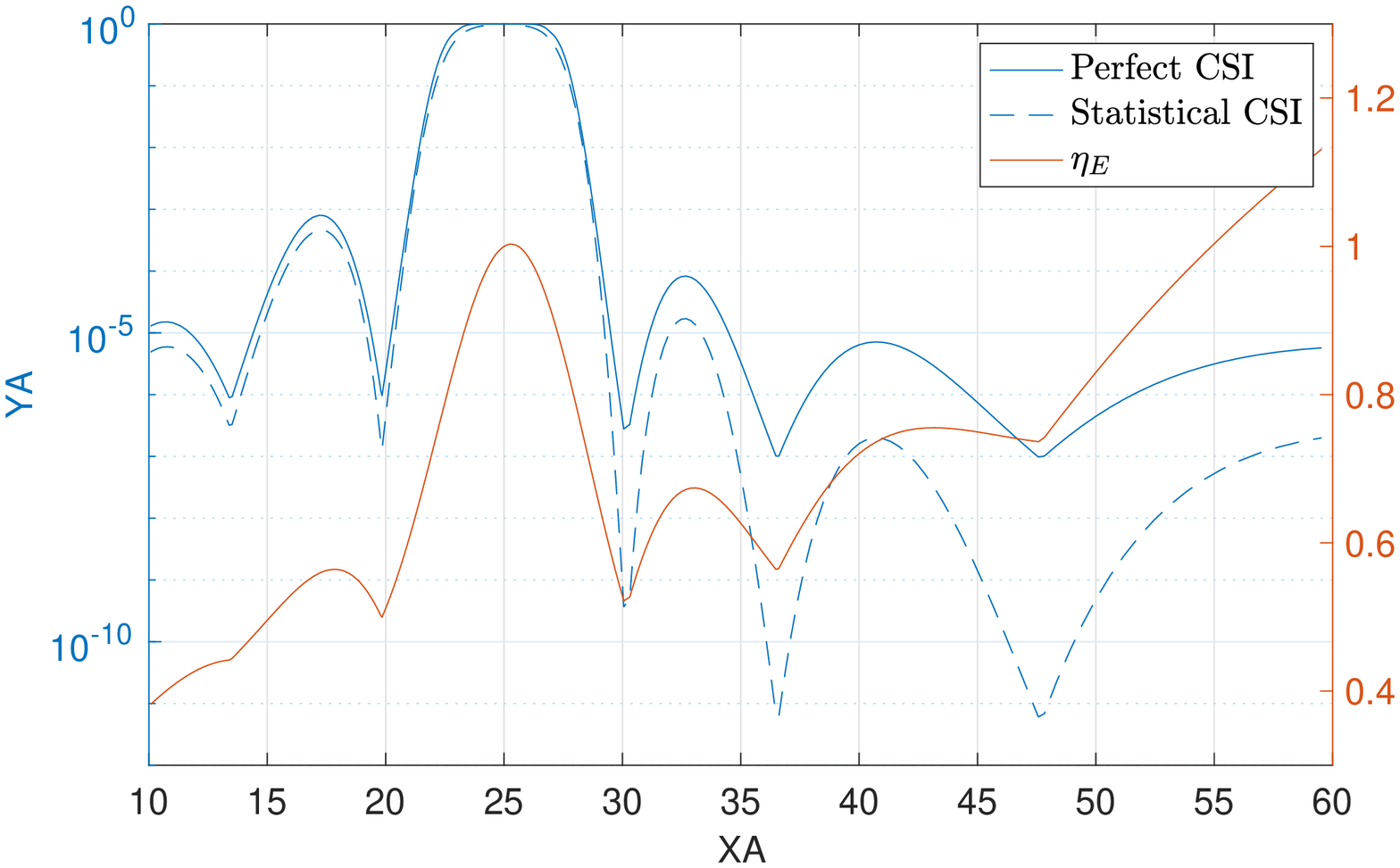}
\caption{}
\label{fig:r_3_1_power_manipulation_attack}
\end{subfigure}%
\begin{subfigure}{0.5\textwidth}
\psfrag{XA}[][]{\small LoS strength $\KRice$ [dB]}
\psfrag{YA}[][]{\small Missed detection prob. $\pMD$}
\includegraphics[width=\columnwidth]{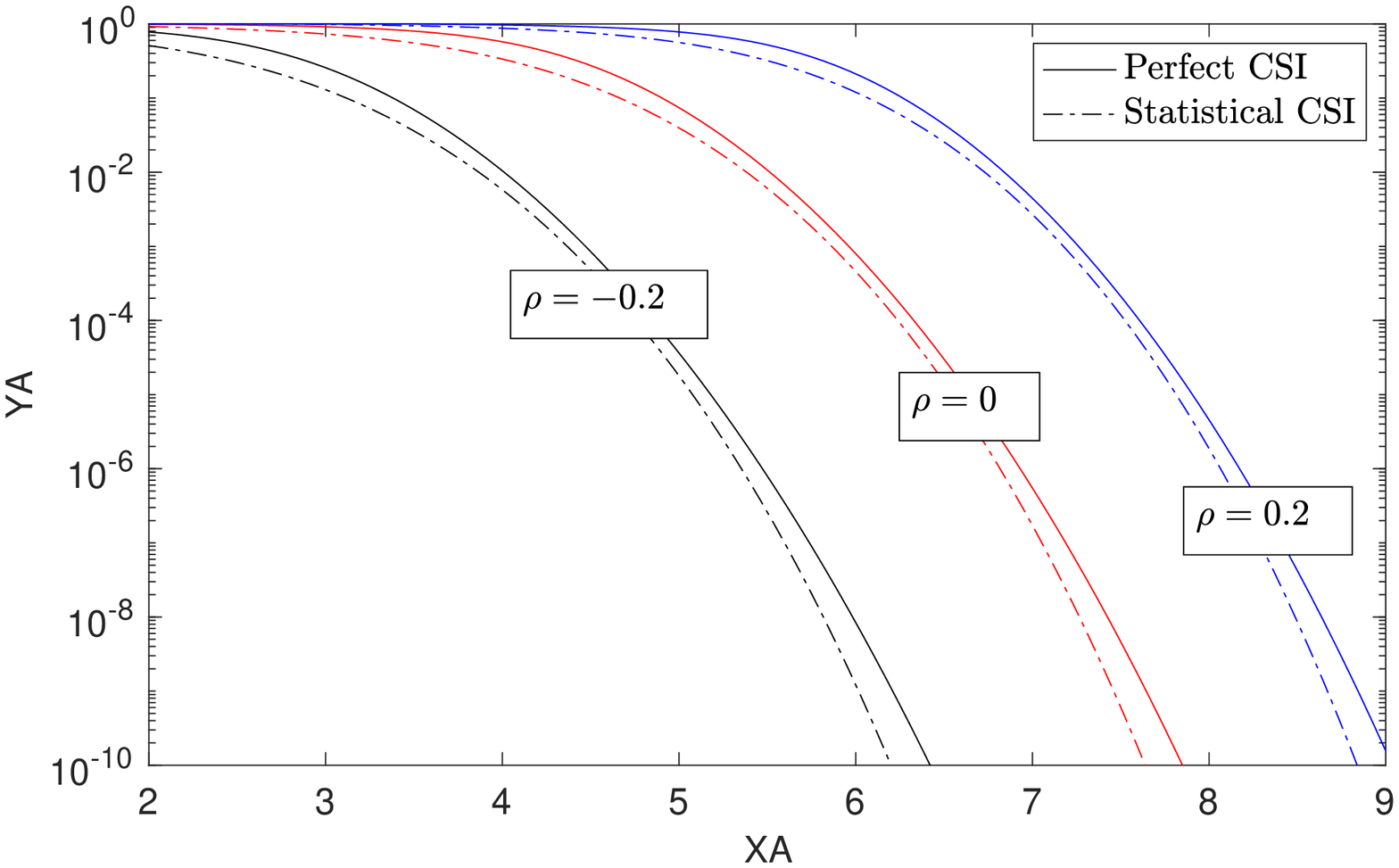}
\caption{}
\label{fig:r_3_2_power_manipulation_KRice}
\end{subfigure} %
\vspace{-2ex}
\caption{Detection performance under power manipulation attack for varying CSI knowledge: (a) for varying attack position and (b) for fixed position with varying Rice factor.}
\vspace{-3ex}
\label{fig:power_manipulation}
\end{figure*}

In Fig.~\ref{fig:power_manipulation}, we illustrate the detection performance under the power manipulation attack with $\Ka=2$ at locations RRH2 and RRH4 and Alice positioned at $\xi_A=(\text{40 m},\text{30 m})$.
In the lefthand axis of Fig.~\ref{fig:r_3_1_power_manipulation_attack}, we show the missed detection probability under the power manipulation strategy for perfect and statistical CSI knowledge at Eve.
The righthand axis shows the corresponding power manipulation amplitude $\eta_E$ for the statistical CSI strategy in~\eqref{eq:fixed_pow_strat}.
We observe that for positions close to RRH4, Eve can achieve close to the optimal success probability with only statistical CSI knowledge.
As expected, we can also observe that the required power manipulation amplitude $\eta_E$ increases with distance since Eve needs to compensate for the higher path-loss compared to Alice's channel.
In Fig.~\ref{fig:r_3_2_power_manipulation_KRice}, we show the same missed detection probabilities but for a fixed attack position $\xi_E=(32\text{ m},30\text{ m})$ and varying LoS strength in terms of the Rice factor $\KRice$.
We again observe that Eve can achieve close to the optimal performance with only statistical CSI knowledge.
However, we also observe that both probabilities decay rapidly with increased LoS strength.

\vspace{-3ex}
\subsection{Validation of Heuristic Search Algorithm for Attack Position Optimization} 
\label{sub:attacker_position}
\definecolor{matlab-blue}{RGB}{39, 113, 189}
\definecolor{matlab-red}{RGB}{218, 83, 24}

\begin{figure*}
\centering
\begin{subfigure}{0.5\textwidth}
\psfrag{XA}[][]{\tiny Distance to RRH4 $d_E^{(1)}$ [m]}
\psfrag{YA}[][][1][-180]{\tiny $\pMD^{(\text{Opt.PMA)})}$}
\psfrag{XB}[][]{\tiny Distance to RRH4 $d_E^{(1)}$ [m]}
\psfrag{YB}[][]{\tiny }
\includegraphics[width=\columnwidth]{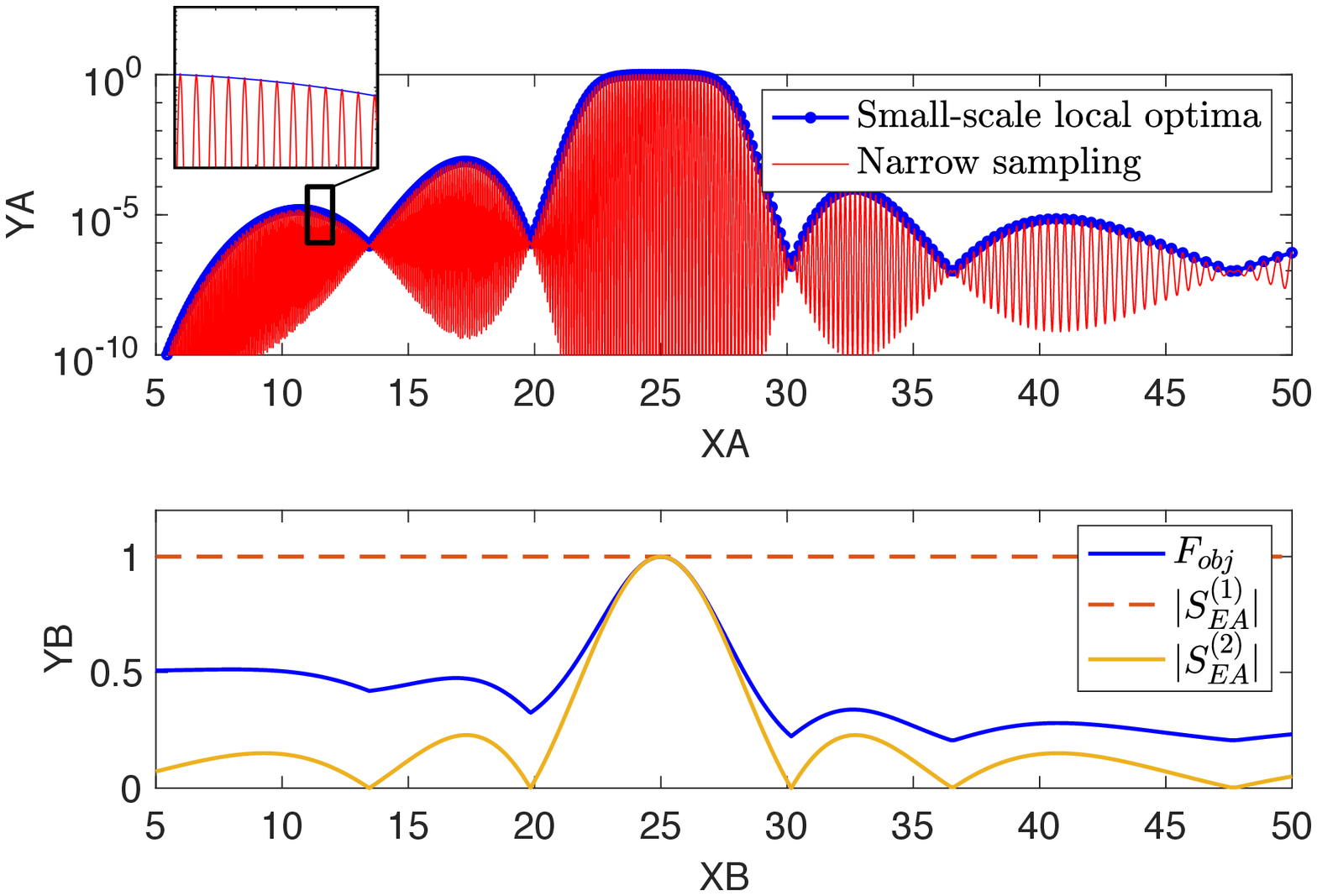}
\caption{}
\label{fig:r_2_optimal_position_line}
\end{subfigure}%
\begin{subfigure}{0.5\textwidth}
\psfrag{X}[][]{\tiny AoA w.r.t RRH4 $\Phi_E^{(1)}$ [rad]}
\psfrag{YA}[][]{\tiny $\pMD^{(wc)}$}
\psfrag{YB}[][]{\tiny}
\includegraphics[width=\columnwidth]{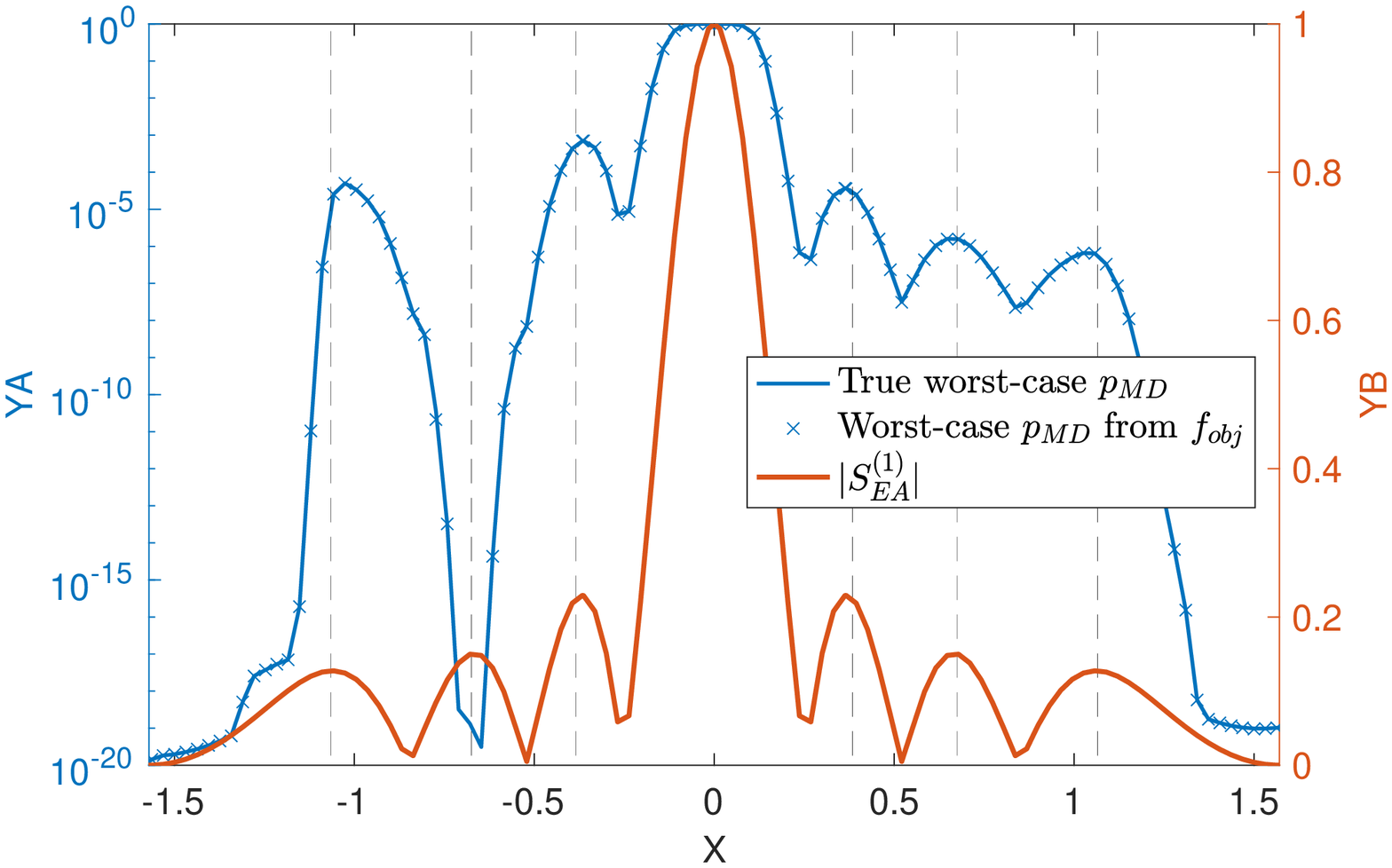}
\caption{}
\label{fig:r_2_optimal_position_circle}
\end{subfigure} %
\vspace{-2ex}
\caption{Missed detection probability under optimal power manipulation for different attack positions: (a) for a line with fixed AoA with respect to RRH4 and (b) for a range of AoAs with respect to RRH4 where distance is optimized to maximize MDP.}
\vspace{-3ex}
\label{fig:small_scale_sampling}
\end{figure*}

In Fig.~\ref{fig:small_scale_sampling} we consider a deployment with $\Ka=2$ arrays with $\NRx=8$ antennas each, positioned according to RRH2 and RRH4 in Fig.~\ref{fig:deployment}.
In this figure the LoS strength is $\KRice=6$ dB.
In Fig.~\ref{fig:r_2_optimal_position_line}, both Alice and Eve are centered in front of RRH4 and Eve is varying her distance with respect to this array.
Firstly, the upper part of Fig.~\ref{fig:r_2_optimal_position_line} illustrates that, by following the result in Lemma \ref{lem:small_scale}, we can appropriately sample the local maxima of $\pMD^{(\text{Opt.PMA)}}$.
We observe that the sampled envelope of the MDP approaches $1-\pFA\approx 1$ when the distance comes close to 25 m since this means that Eve is very close to Alice's position.
The lower part illustrates the objective function $F_{\text{obj}}(\xi_E)$ in \eqref{eq:exp_fobj} and the two angular response inner products $S_{AE}^{(1)}$ and $S_{AE}^{(2)}$ given by \eqref{eq:ang_inn_prod}.
We observe that the envelope of the MDP curve (in the upper plot) closely resembles the shape of the objective function (in the lower plot) which justifies the use of $F_{\text{obj}}$ for optimization in our heuristic approach.
$S_{AE}^{(2)}$ follows the angular response for the second array (i.e., RRH4) and illustrates that the best attack positions on this straight line are located at the intersections with the side lobes of the second array.
The fact that $S_{AE}^{(1)}$ remains constant is expected since Eve is moving along the main lobe of the first array (i.e., RRH3) and does not change the AoA with respect to this array.

In exactly the same scenario as above, Fig.~\ref{fig:r_2_optimal_position_circle} shows the worst-case MDP along a given AoA with respect to RRH4.
This worst-case MDP was obtained by exhaustive search along a straight line for each AoA.
The solid line indicates the true maximum MDP along the corresponding AoA.
The cross markers illustrate the corresponding MDP obtained by maximizing $F_{\text{obj}}(\xi_E)$ along the corresponding AoA.
Their agreement again illustrates that the heuristic approach, that optimizes $F_{\text{obj}}(\xi_E)$, provides valid results.
Moreover, we can observe that if Eve is not allowed close to Alice (e.g., $\Phi_E^{(1)}>0.2$ rad) there is a local optimal AoA around $\Phi_E^{(1)}=0.4$ (the first local maxima for positive $\Phi_E^{(1)}$) that agrees well with the first side lobe of RRH4.

\begin{figure*}[h!]
\begin{subfigure}{0.5\textwidth}
\psfrag{X}[][]{\small $x$ [m]}
\psfrag{Y}[][]{\small $y$ [m]}
\includegraphics[width=\columnwidth]{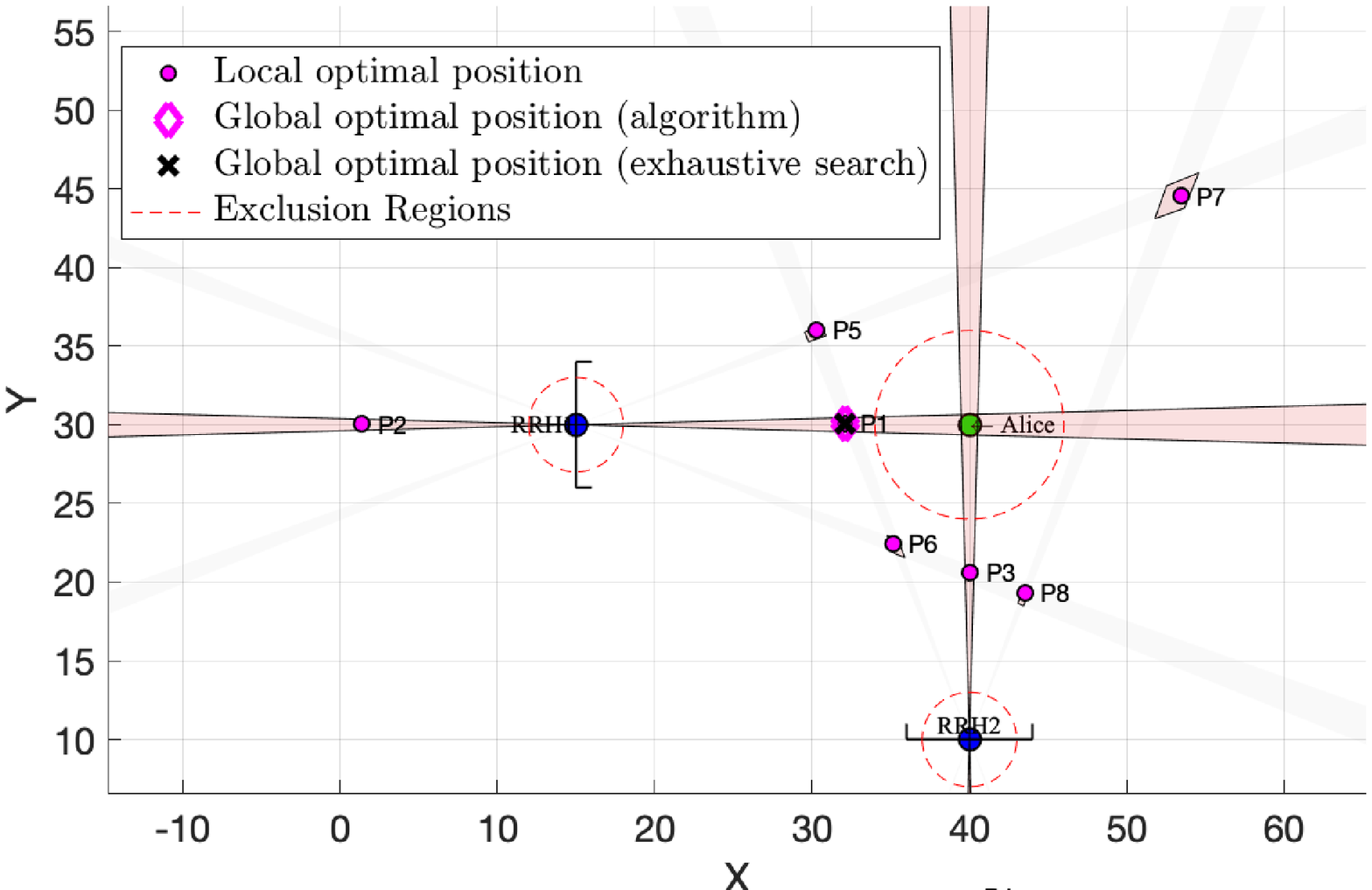}
\caption{}
\label{fig:example_map}
\end{subfigure}%
\begin{subfigure}{0.5\textwidth}
\psfrag{X}[][]{\tiny Position index}
\psfrag{YA}[][]{\tiny \textcolor{matlab-blue}{Objective function $f_{\text{obj}}(\mean_E)$}}
\psfrag{YB}[][][1][-180]{\tiny \textcolor{matlab-red}{Missed detection prob. $\pMD^{(wc)}$}}
\includegraphics[width=\columnwidth]{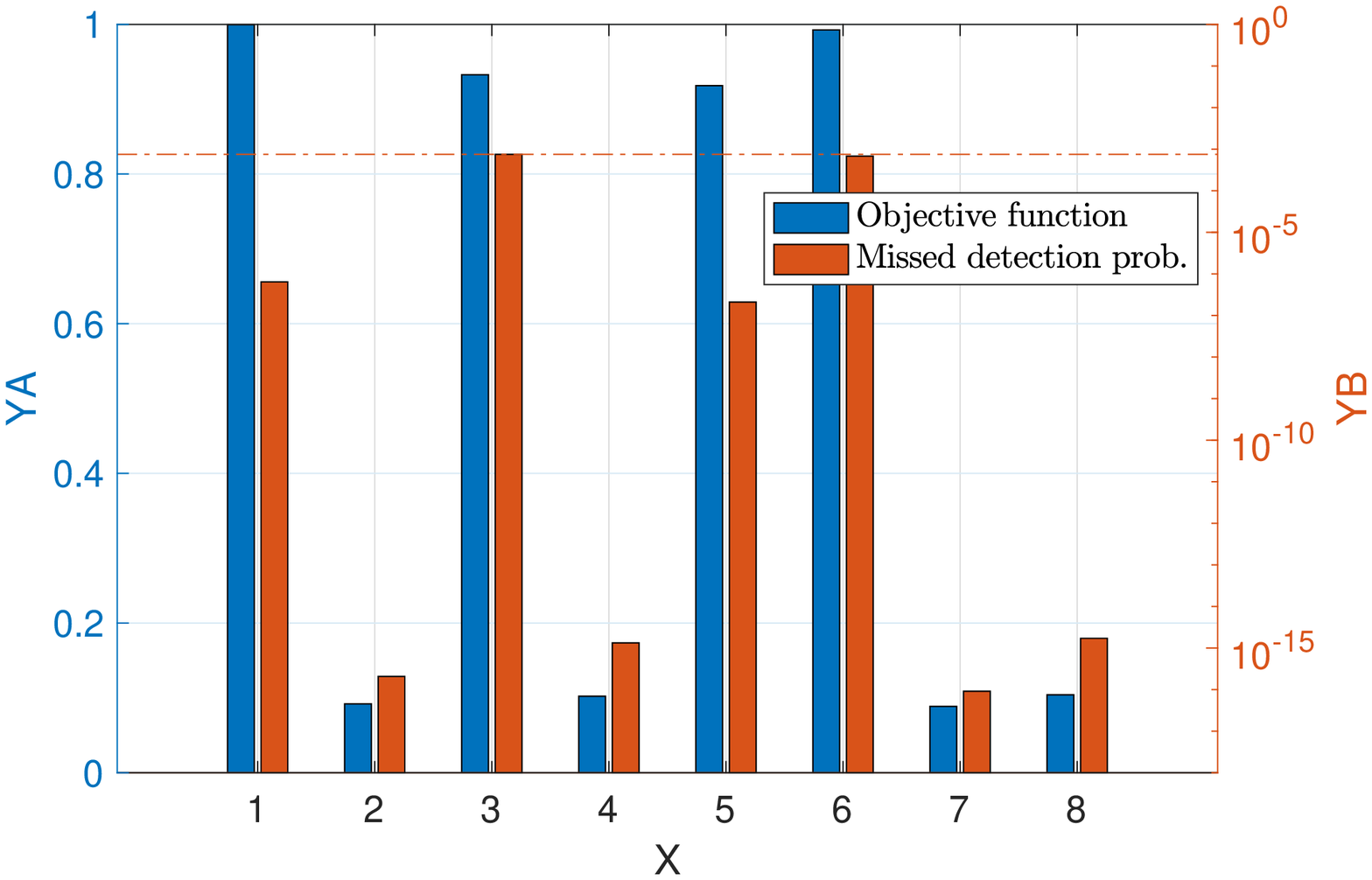}
\caption{}
\label{fig:example_performance}
\end{subfigure} %
\vspace{-2ex}
\caption{Example of optimization for Scenario A: (a) map over considered deployment and marked positions; (b) the corresponding objective function values and MDPs.}
\vspace{-3ex}
\label{fig:optimization_example}
\end{figure*}

Fig.~\ref{fig:optimization_example} exemplifies a result of the optimization algorithm proposed in Section~\ref{sub:heuristic_solution_for_general_rice_fading} over the entire deployment area.
For this example, we have assumed $\KRice = 6$ dB, $\pFA=10^{-2}$, $\rho=0$, $\NRx=8$, and $\Ka=2$.
The red-shaded regions in Fig.~\ref{fig:example_map} mark the areas that are searched, i.e., main lobes and intersections of side lobes.
We have defined the allowed region $\mathcal{R}$ as positions further than $6$ m from Alice and 3 m from the RRHs.
In this map, we show the obtained local optimal positions, and the global optimal position obtained by our approach compared to the true global optimal position obtained by an exhaustive search.
We can see that the heuristic approach finds the true worst-case position that lies within one of the search areas along the main lobe of RRH1.
For this scenario, the worst-case MDP is around $10^{-3}$.
Fig.~\ref{fig:example_performance} shows the objective function values and corresponding MDP for each of the candidate positions.
We observe that the positions with larger objective function values are, in fact, the positions with higher MDP.

\vspace{-3ex}
\subsection{Comparison of Deployment Scenarios}

\begin{figure*}
\centering
 \psfrag{A}[][]{\scriptsize $\Ka=1,\NRx=16$}
 \psfrag{B}[][]{\scriptsize $\Ka=2,\NRx=8$}
 \psfrag{C}[][]{\scriptsize $\Ka=4,\NRx=4$}
 \psfrag{D}[][]{\scriptsize $\Ka=8,\NRx=2$}
\includegraphics[width=\columnwidth]{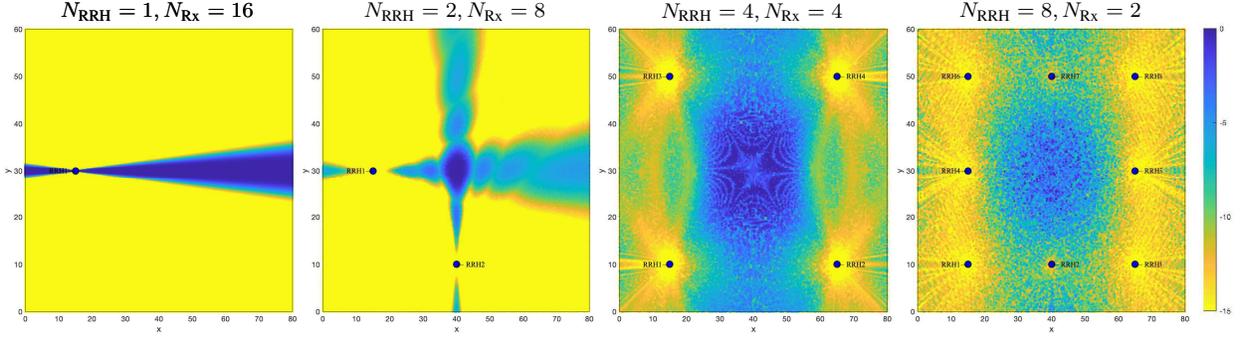}
\vspace{-2ex}
\caption{Heat-maps of log-MDP $\log_{10}(\pMD^{(\text{Opt. PMA})})$ for different deployment scenarios.}
\vspace{-3ex}
\label{fig:heatmap}
\end{figure*}

In Fig.~\ref{fig:heatmap}, we illustrate heat-maps of the log-MDP for four different deployment scenarios.
It is important to note that in every deployment the total number of deployed antennas is fixed to $\Ka\NRx=16$.
We have truncated the MDP values so that the completely yellow regions correspond to a MDP less than $10^{-15}$.
In the second deployment (i.e., with $\Ka=2,\NRx=8$), it is possible to see that many side lobe intersections are associated with slightly increased MDP values indicated by blue regions.

In Table~\ref{tab:1}, we summarize the results from evaluating the search algorithm for different deployment scenarios.
\textcolor{black}{Scenario R is a reference with $\NRx = 16$ antennas deployed at RRH1 (i.e., the same deployment as in the first plot in Fig.~\ref{fig:heatmap}) and Scenario A-G correspond to various configurations of $\NRx$ and $\Na$.}
For these results we applied an exclusion region of points within 6 m from Alice and within 3 m from each RRH.
For each scenario, the table shows the obtained worst-case MDP, the area coverage in percent, number of small-scale sampling positions searched by the algorithm, and the total number of small-scale sampling positions in the entire deployment area.
Area coverage is here defined as the percentage of points with $\pMD^{(\text{Opt. PMA})}<10^{-4}$.

For comparison, consider that an exhaustive search with sampling resolution of $\lambda_c/10=0.0125$ m would result in a complexity on the order of $10^7$ positions in the considered 80 m$\times$60 m area.
From Table~\ref{tab:1}, we see that the final number of search positions are in orders ranging from $10^3$ (Scenario A and D) to $10^5$ (Scenario C), representing a significant complexity reduction compared to the exhaustive search. 
This reduction is generally larger for less distributed scenarios (e.g. only 3.5\% of all the local optima are searched for Scenario A).
The reason for this is twofold: (i) with fewer RRHs, there are fewer main lobes and intersection regions in total, and (ii) with more antennas per array, the beams of each RRH are more directed, and thus, the main lobe and intersection areas become smaller.

\textcolor{black}{The result for the centralized Scenario R was obtained by placing Eve at one of the trivial worst-case positions (i.e., the points resulting in the same AoA as Alice).
It is expected that this results in $\pMD^{(\text{Opt. Position})}=1$ since Eve, given perfect CSI and feature knowledge, always can achieve $d^{(\text{Opt. PMA})} = 0$ at such positions.
The rest of the parameters in Table~\ref{tab:1} are not applicable for Scenario R since the optimization algorithm was not used.}
For the considered settings, the $\Ka=2$ deployments (scenario A and D) provide the lowest worst-case MDP for the given total number of antennas.
These scenarios also result in the highest coverage
We can also observe that the more distributed scenarios $\Ka>2$ all result in a worst-case MDP close to 1.
While noting that this observation depends on the choice of exclusion region, which in this case is a 6 m disc around Alice, these results indicate that distributing antennas to more than $\Ka=2$ locations provides no additional benefits in terms of worst-case missed detection probability.
That is, for a $\Ka=2$ deployment, one should rather invest in adding antennas to the existing RRHs than in distributing the antennas further.

\begin{table*}[]
\begin{tabular}{cccccccc}
Scenario                                       & $\Ka$ & $\NRx$ & Tot. \# antennas & $\pMD^{(\text{Opt. Position})}$ & Coverage & \# Search pos. & \# Small-scale local optima \\ \hline
\rowcolor[HTML]{EFEFEF} 
\multicolumn{1}{c|}{R}                         & 1     & 16      & 16  	&1.000000e-00	&N/A	&N/A&N/A                  \\
\rowcolor[HTML]{EFEFEF} 
\multicolumn{1}{c|}{\cellcolor[HTML]{EFEFEF}A} & 2     & 8      & 16 	&7.469913e-04	&99,96\%	&9730 (3,5\%)	&278263                  
\\
\multicolumn{1}{c|}{B}                         & 4     & 4      & 16  	&9.377572e-01	&97,96\%	&40649 (10,1\%)	&401358                    \\
\rowcolor[HTML]{EFEFEF} 
\multicolumn{1}{c|}{\cellcolor[HTML]{EFEFEF}C} & 8     & 2      & 16 	&6.260576e-01	&99,77\%	&105431 (29,9\%)&352295               	
\\
\multicolumn{1}{c|}{D}                         & 2     & 6      & 12 	&7.367218e-02	&97,84\%	&5923 (2,1\%)	&277892                \\
\rowcolor[HTML]{EFEFEF} 
\multicolumn{1}{c|}{\cellcolor[HTML]{EFEFEF}E} & 3     & 4      & 12 	&9.558150e-01	&94,96\%	&28665 (7,3\%)	&394787              	
\\
\multicolumn{1}{c|}{F}                         & 4     & 3      & 12    &9.803737e-01	&92,23\%	&55750 (13,9\%)	&400514                  \\
\rowcolor[HTML]{EFEFEF} 
\multicolumn{1}{c|}{\cellcolor[HTML]{EFEFEF}G} & 6     & 2      & 12    &9.266438e-01	&96,74\%	&96962 (27,2\%)	&356388               	\\   
\end{tabular}
\vspace{-2ex}
\caption{Summary of results for deployment scenarios A-G and the reference Scenario R.}
\vspace{-5ex}
\label{tab:1}
\end{table*}

\vspace{-3ex}
\subsection{Discussion} 
\label{sub:discussion}

The studied deployment scenarios indicate some lessons on how RRH positioning could be influenced during system design for security purposes.
Firstly, we have observed that a single RRH deployment is very vulnerable to a power manipulation attack as long as the attacker is positioned along the same AoA as the legitimate device (see e.g., first plot in Fig.~\ref{fig:heatmap}).
With that observation in mind, our results clearly illustrate the benefits of a distributed approach compared to the single RRH scenario.
In terms of worst-case MDP, however, our results show no benefits of further distributing the antennas to more than two RRHs, and a comparison of Scenario A and D suggests that narrowing the beams of two existing arrays by adding more antennas is more beneficial than adding additional RRHs.
However, note that this generally is a question of what assumptions we make regarding the exclusion region (i.e., how close to the legitimate device can an attacker realistically be expected to come).
Compare, for example, the deployment $\Ka=2$ and $\Ka=8$ in Fig.~\ref{fig:heatmap} where it can be argued that the 8 RRH deployment would be preferable to protect against attackers close to the edges of the deployment area (e.g., in a case where it is unlikely that an attacker can get physically close to the legitimate device).
To fully answer such questions, further assumptions on the type of system are needed like, e.g., if the system is publicly deployed, if devices are mobile or stationary, and if it is likely that an internal device could be an attacker, etc.

\textcolor{black}{
Making a numerical comparison of our obtained results to previous PLA schemes is complicated since the performance often is highly dependent on channel parameters like, e.g., frequency selectivity, bandwidth, and signal-to-noise ratio.
However, considering the clear benefits from the distributed approach, we can conjecture how these results would extend to other PLA schemes and feature choices (e.g., frequency or impulse response-based schemes~\cite{Xiao2007,Liu2011}):
Assuming a fair parameterization  and  scaling  of features  (e.g.,  number  of  independent  resolvable channel taps of the impulse response matches the number of antennas per array; same SNR), we expect that the performance of GLRT based schemes that employ other features will coincide with the single-RRH case.
This implies that previous PLA approaches also would benefit from combining features from multiple distributed receivers.
For example, the system model considered in this paper could be extended to a multi-carrier system (i.e., stacking the feature-vector corresponding to each sub-carrier of an OFDM channel), which would introduce an additional feature dimension and effectively result in a distributed version of the scheme considered in~\cite{Xiao2007}.
The methods from Section~\ref{sec:opt_power_attack} generalize to this extended system model since the feature vector remains a CSCG vector.
Hence, the theoretical results obtained in this paper could be used to evaluate the performance improvements from this combined approach, which we expect would depend on the correlation across sub-carriers and antennas (i.e., smaller correlation would correspond to greater performance improvements).
}

Finally, the results also provide some insights into how detection performance depends on the channel model parameters.
Channel characteristics such as LoS strength are obviously mostly determined by the environment where the system is deployed.
Here, we can only conclude that the studied authentication scheme would be better suitable where strong LoS channels are expected (e.g., in an open industry hall or a road side).
Fading correlation, which according to our results improves detection performance, can to some extent be influenced through system design by densely spaced arrays; however, note that in our model such a modification would also affect the normalized antenna separation, and consequently, the beam-forming pattern.

\vspace{-2ex}
\section{Conclusion}		\label{sec:conclusion}
In this paper, we have presented mathematical tools for analyzing the worst-case detection performance of a physical layer authentication scheme subject to an attacker that uses the optimal transmit power and phase and the optimal attack position.
Such bounds are highly relevant for the application of feature-based PLA in systems that require strict security guarantees, as exemplified by mission-critical communications and URLLC.
In particular, the worst-case bounds have been derived for a distributed PLA approach that fits well within the distributed MIMO architectures considered for many mission-critical applications and 5G radio-access networks.
Our results indicate that a distributed architecture can significantly reduce the worst-case missed detection probability for fixed number of antennas.
However, our worst-case analysis has shown no performance improvements from distributing antennas on more than two RRHs.
That is, for a distributed system with two arrays, instead of adding a third array one should rather use additional antennas to narrow the beams of the existing two arrays. 

{
\footnotesize
\bibliography{phd_bibliography.bib}

\begin{thebibliography}{10}
\providecommand{\url}[1]{#1}
\csname url@samestyle\endcsname
\providecommand{\newblock}{\relax}
\providecommand{\bibinfo}[2]{#2}
\providecommand{\BIBentrySTDinterwordspacing}{\spaceskip=0pt\relax}
\providecommand{\BIBentryALTinterwordstretchfactor}{4}
\providecommand{\BIBentryALTinterwordspacing}{\spaceskip=\fontdimen2\font plus
\BIBentryALTinterwordstretchfactor\fontdimen3\font minus
  \fontdimen4\font\relax}
\providecommand{\BIBforeignlanguage}[2]{{%
\expandafter\ifx\csname l@#1\endcsname\relax
\typeout{** WARNING: IEEEtran.bst: No hyphenation pattern has been}%
\typeout{** loaded for the language `#1'. Using the pattern for}%
\typeout{** the default language instead.}%
\else
\language=\csname l@#1\endcsname
\fi
#2}}
\providecommand{\BIBdecl}{\relax}
\BIBdecl

\bibitem{Hou2014}
W.~Hou, X.~Wang, J.-Y. Chouinard, and A.~Refaey, ``Physical layer
  authentication for mobile systems with time-varying carrier frequency
  offsets,'' \emph{IEEE Transactions on Communications}, vol.~62, no.~5, pp.
  1658--1667, May 2014.

\bibitem{Jana2010}
S.~Jana and S.~K. Kasera, ``On fast and accurate detection of unauthorized
  wireless access points using clock skews,'' \emph{IEEE Transactions on Mobile
  Computing}, vol.~9, no.~3, pp. 449--462, March 2010.

\bibitem{Danev2009}
B.~Danev and S.~Capkun, ``Transient-based identification of wireless sensor
  nodes,'' in \emph{2009 International Conference on Information Processing in
  Sensor Networks}, April 2009, pp. 25--36.

\bibitem{Hussain2009}
S.~Hussain, ``Using received signal strength indicator to detect node
  replacement and replication attacks in wireless sensor networks,'' in
  \emph{Proceedings of SPIE - The International Society for Optical
  Engineering}, vol. 7344, 04 2009.

\bibitem{Xiao2007}
L.~Xiao, L.~Greenstein, N.~Mandayam, and W.~Trappe, ``Fingerprints in the
  ether: Using the physical layer for wireless authentication,'' in \emph{IEEE
  Intl. Conference on Communications}, June 2007, pp. 4646--4651.

\bibitem{Abdelaziz2019}
A.~{Abdelaziz}, C.~E. {Koksal}, F.~{Barickman}, R.~{Burton}, J.~{Martin}, and
  J.~{Weston}, ``Enhanced authentication based on angle of signal arrivals,''
  \emph{IEEE Transactions on Vehicular Technology}, vol.~68, no.~5, pp. 1--1,
  2019.

\bibitem{Baracca2012}
P.~Baracca, N.~Laurenti, and S.~Tomasin, ``Physical layer authentication over
  {MIMO} fading wiretap channels,'' \emph{IEEE Transactions on Wireless
  Communications}, vol.~11, no.~7, pp. 2564--2573, July 2012.

\bibitem{Yu2008}
P.~L. Yu, J.~S. Baras, and B.~M. Sadler, ``Physical-layer authentication,''
  \emph{IEEE Transactions on Information Forensics and Security}, vol.~3,
  no.~1, pp. 38--51, March 2008.

\bibitem{Chen2019}
R.~{Chen}, C.~{Li}, S.~{Yan}, R.~{Malaney}, and J.~{Yuan}, ``Physical layer
  security for ultra-reliable and low-latency communications,'' \emph{IEEE
  Wireless Communications}, vol.~26, no.~5, pp. 6--11, October 2019.

\bibitem{Weinand2019}
A.~{Weinand}, R.~{Sattiraju}, M.~{Karrenbauer}, and H.~D. {Schotten},
  ``Supervised learning for physical layer based message authentication in
  {URLLC} scenarios,'' in \emph{IEEE Vehicular Technology Conference}, Sep.
  2019, pp. 1--7.

\bibitem{Panigrahi2017}
S.~R. {Panigrahi}, N.~{Bjorsell}, and M.~{Bengtsson}, ``Feasibility of large
  antenna arrays towards low latency ultra reliable communication,'' in
  \emph{IEEE International Conference on Industrial Technology}, 2017, pp.
  1289--1294.

\bibitem{Alonzo2020}
M.~{Alonzo}, P.~{Baracca}, S.~R. {Khosravirad}, and S.~{Buzzi}, ``{URLLC} for
  factory automation: an extensive throughput-reliability analysis of
  {D-MIMO},'' in \emph{ITG Workshop on Smart Antennas}, 2020, pp. 1--6.

\bibitem{Ferrante2015}
A.~{Ferrante}, N.~{Laurenti}, C.~{Masiero}, M.~{Pavon}, and S.~{Tomasin}, ``On
  the error region for channel estimation-based physical layer authentication
  over rayleigh fading,'' \emph{IEEE Transactions on Information Forensics and
  Security}, vol.~10, no.~5, pp. 941--952, 2015.

\bibitem{Senigagliesi2020}
\BIBentryALTinterwordspacing
L.~Senigagliesi, M.~Baldi, and E.~Gambi, ``Performance of statistical and
  machine learning techniques for physical layer authentication,'' \emph{CoRR},
  Jan 2020. [Online]. Available: \url{https://arxiv.org/abs/2001.06238}
\BIBentrySTDinterwordspacing

\bibitem{Xiao2008}
L.~Xiao, L.~Greenstein, N.~Mandayam, and W.~Trappe, ``Using the physical layer
  for wireless authentication in time-variant channels,'' \emph{IEEE
  Transactions on Wireless Communications}, vol.~7, no.~7, pp. 2571--2579, July
  2008.

\bibitem{Abdelaziz2016}
\BIBentryALTinterwordspacing
A.~Abdelaziz, R.~Burton, and C.~E. Koksal, ``Message authentication and secret
  key agreement in {VANET}s via angle of arrival,'' \emph{CoRR}, Sep. 2016.
  [Online]. Available: \url{http://arxiv.org/abs/1609.03109}
\BIBentrySTDinterwordspacing

\bibitem{Mahmood2017}
A.~Mahmood, W.~Aman, M.~O. Iqbal, M.~M.~U. Rahman, and Q.~H. Abbasi, ``Channel
  impulse response-based distributed physical layer authentication,'' in
  \emph{IEEE Vehicular Technology Conference}, June 2017, pp. 1--5.

\bibitem{Forssell2019b}
H.~Forssell, R.~Thobaben, and J.~Gross, ``Performance analysis of distributed
  simo physical layer authentication,'' in \emph{IEEE International Conference
  on Communications}, May 2019, pp. 1--6.

\bibitem{Yan2014}
S.~{Yan} and R.~{Malaney}, ``Line-of-sight based beamforming for security
  enhancements in wiretap channels,'' in \emph{International Conference on IT
  Convergence and Security}, 2014, pp. 1--4.

\bibitem{Forssell2019}
H.~{Forssell}, R.~{Thobaben}, H.~{Al-Zubaidy}, and J.~{Gross}, ``Physical layer
  authentication in mission-critical {MTC} networks: A security and delay
  performance analysis,'' \emph{IEEE Journal on Selected Areas in
  Communications}, vol.~37, no.~4, pp. 795--808, April 2019.

\bibitem{MacLeod2005}
H.~MacLeod, C.~Loadman, and Z.~Chen, ``Experimental studies of the 2.4-{GHz}
  {ISM} wireless indoor channel,'' in \emph{3rd Annual Communication Networks
  and Services Research Conference}, May 2005, pp. 63--68.

\bibitem{Naffouri2016}
T.~Y. {Al-Naffouri}, M.~{Moinuddin}, N.~{Ajeeb}, B.~{Hassibi}, and A.~L.
  {Moustakas}, ``On the distribution of indefinite quadratic forms in gaussian
  random variables,'' \emph{IEEE Transactions on Communications}, vol.~64,
  no.~1, pp. 153--165, Jan 2016.

\bibitem{Liu2011}
F.~J. Liu, X.~Wang, and H.~Tang, ``Robust physical layer authentication using
  inherent properties of channel impulse response,'' in \emph{Military
  Communications Conference}, Nov 2011, pp. 538--542.

\end{thebibliography}
}

\appendix

\subsection{Optimal Power Manipulation Strategy}
\label{app:optimal_power}

Recall that $d(\rho_E e^{j\varphi_E}\h_E)$ is the discriminant function that the attacker is trying to minimize.
We note that by using a Cholesky factorization $\Cov_A^{-1} = \Q_A^\dag \Q_A$, we can write

\begin{equation}
\begin{aligned}
d(\rho_E e^{j\varphi_E}\h_E) 
&=\QF[\Cov_A^{-1}]{\rho_E e^{j\varphi_E}\h_E-\mean_A}
= 2 \|\Q_A(\rho_E e^{j\varphi_E}\h_E-\mean_A) \|^2 \\
&= 2( \rho_E^2\|\Q_A\h_E\|^2 + \|\Q_A\mean_A\|^2
-2\rho_E\Re\{e^{j\varphi_E} (\Q_A\mean_A)^\dag\Q_A\h_E \}).
\end{aligned}
\label{eq:d_euclidean}
\end{equation}
Taking the derivative w.r.t. $\rho_E$, we get $\frac{\partial }{\partial \rho_E}d(\rho_E e^{j\varphi_E}\h_E)= 4\rho_E\|\Q_A\h_E\|^2 - 4\Re\{e^{j\varphi_E} (\Q_A\mean_A)^\dag\Q_A\h_E \}$
%
and, hence, $\rho_E^*$ such that $\frac{\partial }{\partial \rho_E}d(\rho_E e^{j\varphi_E}\h_E) =0$ results in

\begin{equation}
\rho_E^* = \frac{\Re\{e^{j\varphi_E} (\Q_A\mean_A)^\dag\Q_A\h_E \}}{\|\Q_A\h_E\|^2} =  \frac{\Re\{e^{j\varphi_E}\mean_A^\dag\Cov_A^{-1}\h_E \}}{\h_E^\dag\Cov_A^{-1}\h_E}.
\label{eq:rho_opt}
\end{equation}
Plugging \eqref{eq:rho_opt} into \eqref{eq:d_euclidean}, we get 
\begin{equation}
d(\rho_E^* e^{j\varphi_E}\h_E) = 2\left(\mean_A^\dag\Cov_A^{-1}\mean_A - \frac{\Re\{e^{j\varphi_E}\mean_A^\dag\Cov_A^{-1}\h_E \}^2}{\h_E^\dag\Cov_A^{-1}\h_E}\right).
\label{eq:d_euclidean2}
\end{equation}
Clearly, $d(\rho_E^* e^{j\varphi_E}\h_E)$ is minimized when $e^{j\varphi_E}\mean_A^\dag\Cov_A^{-1}\h_E$ is real-valued so that $\Re\{e^{j\varphi_E}\mean_A^\dag\Cov_A^{-1}\h_E \}=|\mean_A^\dag\Cov_A^{-1}\h_E| $, which is obtained by setting $\varphi_E^* = -\arg\{\mean_A^\dag\Cov_A^{-1}\h_E\}$. Plugging $\varphi_E^*$ into \eqref{eq:d_euclidean2} and rearranging yields \eqref{eq:dmin} which completes the proof.

\subsection{Integral Representation}
\label{app:int_rep}

Introducing, $\bar{\h}_E = \b + \h$, we can integrate over the PDF of $\h\sim\CN(\mathbf{0},\mathbf{I})$ according to

\begin{equation}
\begin{aligned}
\pMD^{(\text{wc})}(T) = \int_{-\infty}^{\infty} \frac{1}{\pi^N}e^{-\h^\dag\h}\mathcal{U}((\b+\h)^\dag\mathbf{A}(\b+\h))d\h, 
\end{aligned}
\label{eq:gaussIntegral}
\end{equation}
where $\mathcal{U}(\cdot)$ denotes the Heaviside step function. 
By using the Laplace representation\footnote{According to \cite{Naffouri2016}, it is more convenient to work with $1-\mathcal{U}(x)$ when evaluating the complementary CDF.} $1-\mathcal{U}(x)= -\frac{1}{2\pi}\int_{-\infty}^{\infty}\frac{e^{-x(j\omega-\beta)}}{j\omega-\beta}d\omega$ valid for $\beta>0$, we can rewrite \eqref{eq:gaussIntegral} as 

\begin{equation}
\begin{aligned}
\int_{-\infty}^{\infty} \frac{1}{\pi^N}e^{-\h^\dag\h}
\left[-\frac{1}{2\pi}\int_{-\infty}^{\infty}\frac{e^{-(j\omega-\beta)(\b+\h)^\dag\mathbf{A}(\b+\h)}}{j\omega-\beta}d\omega\right] d\h
= -\frac{1}{2\pi^{N+1}}\Int_{-\infty}^{\infty}\Int_{-\infty}^{\infty}\frac{e^{-\h^\dag\h-(\b+\h)^\dag (j\omega-\beta)\mathbf{A}(\b+\h)}}{j\omega-\beta}d\omega d\h.
\end{aligned}
\label{eq:rrr}
\end{equation}
Using the decomposition $\mathbf{A}=\U\D\U^\dag$, the transformations $\tilde{\h} = \U^\dag\h$ and $\bar{\b} = \U^\dag\b$, and the fract that $d\h = d\tilde{\h}$, we can write \eqref{eq:rrr} as

\begin{equation}
\begin{aligned}
-\frac{1}{2\pi^{N+1}}\Int_{-\infty}^{\infty}\Int_{-\infty}^{\infty}\frac{e^{-\tilde{\h}^\dag\tilde{\h}-(\tilde{\h} + \bar{\b})^\dag (j\omega-\beta)\D(\tilde{\h} + \bar{\b})}}{j\omega-\beta}d\omega d\tilde{\h}
= -\frac{1}{2\pi^{N+1}}\Int_{-\infty}^{\infty}\Int_{-\infty}^{\infty}\frac{e^{-(\tilde{\h} + \tilde{\b})^\dag\mathbf{B}(\tilde{\h} + \tilde{\b}) -c(\omega)}}{j\omega-\beta}d\omega d\tilde{\h},
\end{aligned}
\label{eq:aaa}
\end{equation}
with $\mathbf{B} = \mathbf{I}+(j\omega-\beta)\D$, 
$\tilde{\b} = \left(\mathbf{I} + \frac{1}{j\omega-\beta}\D^{-1}\right)^{-1}\bar{\b}$, and 
$c(\omega) = \bar{\b}^{\dag}\left(\mathbf{I} + \frac{1}{j\omega-\beta}\D^{-1}\right)^{-1}\bar{\b}$.
We can then integrate out $\tilde{\h}$ by noting that \eqref{eq:aaa} can be written as

\begin{equation}
\begin{aligned}
-\frac{1}{2\pi}\Int_{-\infty}^{\infty}\Int_{-\infty}^{\infty} \frac{1}{\pi^N} e^{-(\tilde{\h} + \tilde{\b})^\dag\mathbf{B}(\tilde{\h} + \tilde{\b})} d\tilde{\h}\frac{e^{-c(\omega)}}{j\omega-\beta}d\omega
=-\frac{1}{2\pi}\int_{-\infty}^{\infty}\frac{e^{-c(\omega)}}{(j\omega-\beta)|\mathbf{I}+(j\omega-\beta)\D|}d\omega,
\end{aligned}
\end{equation}
where we have used the Gaussian integral solution $\int_{-\infty}^{\infty} \frac{1}{\pi^N} e^{-(\tilde{\h} + \tilde{\b})^\dag\mathbf{B}(\tilde{\h} + \tilde{\b})} d\tilde{\h} = \frac{1}{|\mathbf{B}|}$.




%
%

\end{document}